\documentclass[%
 reprint,
floatfix,
superscriptaddress,
shofootinbib,
 amsmath,amssymb,
 aps,
pra,
]{revtex4-2}
\DeclareUnicodeCharacter{2212}{-}
\usepackage{xcolor}
\usepackage{times}
\usepackage{tikz}
\usepackage{braket}
\usepackage{amssymb}
\usepackage{comment}
\usepackage{dsfont}
\usepackage{graphicx}
\usepackage{dcolumn}
\usepackage{bm}
\usepackage{epstopdf}
\usepackage[colorlinks]{hyperref}
\hypersetup{colorlinks = true,
linkcolor = blue,
filecolor = blue,
urlcolor = blue,
citecolor = blue}

\usepackage[left]{lineno} 

\begin{document}
\title{Topological charge excitations and Green's function zeros in paramagnetic Mott insulators.}
\author{Emile Pangburn}
\affiliation{Universit\'{e} Paris-Saclay, Institut de Physique Th\'eorique,  CEA, CNRS, F-91191 Gif-sur-Yvette, France}

\author{Catherine P\'epin}
\affiliation{Universit\'{e} Paris-Saclay, Institut de Physique Th\'eorique,  CEA, CNRS, F-91191 Gif-sur-Yvette, France}

\author{Anurag Banerjee}
\affiliation{Universit\'{e} Paris-Saclay, Institut de Physique Th\'eorique,  CEA, CNRS, F-91191 Gif-sur-Yvette, France}

\begin{abstract}
We investigate the emergence of topological features in the charge excitations of Mott insulators in the Chern-Hubbard model. In the strong correlation regime, treating electrons as the sum of holons and doublons excitations, we compute the topological phase diagram of Mott insulators at half-filling using composite operator formalism. The Green function zeros manifest as the tightly bound pairs of such elementary excitations of the Mott insulators. Our analysis examines the winding number associated with the occupied Hubbard bands and the band of Green's function zeros. We show that both the poles and zeros show gapless states and zeros, respectively, in line with bulk-boundary correspondence. The gapless edge states emerge in a junction geometry connecting a topological Mott band insulator and a topological Mott zeros phase. These include an edge electronic state that carries a charge and a charge-neutral gapless zero mode. Our study is relevant to several twisted materials with flat bands where interactions play a dominant role.
\end{abstract}
\maketitle
\section{Introduction\label{introduction}}
Band topology in single-electron systems relies on Bloch eigenstates to identify and classify topological properties~\cite{Haldane,KaneMele,bernevig2006quantum,FuKM,MooreBalents,AltlandZirnbauer,TISC_RMP,ColloquiumTI, bradlyn2017topological,kruthoff2017topological}. In correlated systems, however, the key topological features reflect in the Green's functions. While the poles of two-point Green's functions, representing quasiparticles, have been the primary focus, strongly correlated electrons also harbor Green's function zeros (GFZ). These zeros occur in frequency-momentum space where self-energy diverges~\cite{Dzyaloshinskii,KaneZeros} and indicate the presence of composite particles in high-energy physics~\cite{Weinberg}. GFZ has recently attracted considerable interest from the condensed matter community due to its prospect of exhibiting topological properties in strongly correlated systems. The Hatsugai-Kohomoto model~\cite{HK} has been pivotal in investigating the origins of such topological zeros in Mott insulators~\cite{setty2023electronic,setty2023symmetry}, where holons and doublons emerge as elementary excitations of electrons, enabling the exact solvability of the system~\cite{Phillips2020}.

This work offers an innovative and practical pathway for exploring topology in strongly correlated electrons in general. The central idea is to identify the elementary excitations of the strongly correlated ground state and describing them as nearly free quasiparticles. In this paper, Mottness arises from splitting electrons into holons and doublons, which we treat as fundamental excitations. By reviving the composite operator method (COM)~\cite{hubbard1963electron,Beenen0,avella1998hubbard,stanescu2000d,avella2011composite,Haurie_2024}, we provide a plain-vanilla model for the anomalous quantum Hall state in Mott insulators. This framework integrates the study of the topology of charge excitations, Green's function zeros, and their interactions through edge states in a strong interaction regime. Remarkably, Green's function zeros emerge naturally in this framework as tightly bound holon-doublon pairs, thus presenting compelling insights into Mott physics.


Recently, Moir\'e materials have emerged as a promising platform for realizing band topology in strongly correlated electronic systems~\cite{Li2021,SerlinExp,Chen2020}. Such phenomenon can be modeled as transitions from trivial Mott insulators to anomalous quantum Hall insulators in the Bernevig–Hughes–Zhang (BHZ) model~\cite{YoshidaBHZ,AmaricciBHZ,Hohenadler_2013,TadaBHZ}, Kane-Mele model~\cite{KMHubb0,KMHubb1,KMHubb2,bercx2014kane}, and Haldane model~\cite{LeeBHZ,vanhala2016topological,tupitsyn2019phase,he2024phase}, with generic Hubbard interactions at half-filling or quarter filling~\cite{Mai2023}. Furthermore, Green's function zeros are recently studied in the two orbital Hubbard model using the continued fraction expansion~\cite {wagner2023Mott}, slave-boson formalism~\cite{wagner2023edge}. Remarkably, bands of GFZ can acquire a topological character with equivalent bulk-boundary correspondence~\cite{wagner2023Mott}, which manifests as gapless zero modes. 

This paper examines the two-band Chern model in square lattice~\cite{hughes2011inversion} with strong on-site repulsion. Our key finding identifies a parameter range where the occupied single-particle excitation bands acquire a non-zero Chern number. Additionally, we show that the interacting two-point Green's function can become topological, supporting edge zero modes without closing the single-particle excitation gap. We also analyze a junction between a topological Mott band insulator (TMBI) and a topological Mott zero state (TMZ) with non-zero winding numbers, indicating the presence of gapless zeros and poles. Our findings can be relevant for understanding the quantum anomalous Hall effect observed in twisted bilayer systems~\cite{Li2021,SerlinExp,Chen2020,Tang2020}.

\section{Model and method}
We work with the two-band Chern-Hubbard model
\begin{align}
\mathcal{H}=& M\sum\limits_{i,\sigma} c^\dagger_{i\alpha\sigma}  \left[\hat{\tau}_z\right]_{\alpha\alpha^\prime} c_{i\alpha^\prime\sigma}+\sum\limits_{\langle i,j \rangle ,\alpha,\alpha^\prime\sigma} c^\dagger_{j\alpha \sigma} \left[\hat{t}_{ij}\right]_{\alpha \alpha^\prime}  c_{i \alpha^\prime \sigma}  \nonumber \\
& -\mu \sum_{i\alpha\sigma} \hat{n}_{i\alpha\sigma}+ \sum \limits_{i \alpha} U_\alpha \hat{n}_{i \alpha \uparrow} \hat{n}_{i \alpha \downarrow} 
\label{eq:Hamiltonian}
\end{align} 
Here, $c_{i\alpha\sigma}$ $(c^\dagger_{i\alpha\sigma})$ annihilates (creates) an electron at site $i$ with orbital $\alpha$ and spin $\sigma$, where $\sigma=\uparrow,\downarrow$ for spin-$1/2$ electrons and ${\alpha=s,p}$ on a square lattice in two dimensions (2D). The nearest neighbor hoppings are allowed. The $M$ is a crystal field spiltting term and acts with an opposite sign on two orbitals. The second term is nearest neighbor hopping between the nearest neighbour sites such that for the square lattice is $\delta=\pm \hat{x}, \pm \hat{y}$.  At each site $j$ the inter and intraorbital hoppings are  given by the $2\times2$ matrix,
\begin{align}
    \hat{t}_{j, j\pm\delta}=t\left( \hat{\tau}_z+ i  \hat{\tau}_{\delta}\right)
\end{align}
All  nearest neighbor interorbital and intraorbital hopping are of strength given by $t$ and all energy scales are in units of $t=1$. Furthermore, $\hat{\tau}_\alpha$ are the different components of the Pauli Matrices that acts in the orbital degrees of freedom. The hopping remains spin-independent. For $U=0$, the Hamiltonian is exactly a $2$-band square lattice Chern model~\cite{qi2008topological}.

The $U$ term is an on-site repulsion between electrons on the same orbitals. The number operator is defined as $\hat{n}_{i\alpha\sigma}=c^\dagger_{i\alpha\sigma}c_{i\alpha\sigma}$, and the local electron density is the expectation value of the number operator,
$n_{i\alpha\sigma} = \langle\hat{n}_{i\alpha\sigma}\rangle$. The chemical potential $\mu$ fixes the total electron density of the system $n=(1/N)\sum_{i,\alpha,\sigma} \langle \hat{n}_{i\alpha\sigma} \rangle$ which can vary from $n \in [0,4]$. Here $N$ is is the total number of sites in the system. Also we defined average orbital electron density as $n_\alpha=(1/N)\sum_{i,\sigma} \langle \hat{n}_{i,\alpha,\sigma} \rangle$.  In this study, we neglect any intra-orbital long range interactions and inter-orbital interactions.

\subsection{Composite operators method}
The composite operator method (COM)~\cite{avella2011composite} has been recently reviewed for two-bands~\cite{pangburn2024spontaneous} and real space~\cite{CDW_COM_AB}. The COM approach involves identifying a set of operators, in the atomic limit ($t=0$) that represent the key excitations from a singly occupied ground state. In the following we consider the holon $\xi_{i\alpha\sigma}=c_{i\alpha\sigma}\left(1-n_{i\alpha\overline{\sigma}}\right)$ and doublon $\eta_{i\alpha\sigma}=c_{i\alpha\sigma}n_{i\alpha\overline{\sigma}}$.

The method relies on using equation of motion of composite Green's function, to include hopping terms. Note that in this formalism the electrons splits into the holon and doublons exciatations such that $c=\xi+\eta$ and hence single particle electronic Green's function can be constructed from the composite Green's function.  In the following, we use the paramagnetic basis at each site
$\mathbf{\Psi}_i=\left(\xi_{is\sigma}, \eta_{is\sigma}, \xi_{ip\sigma}, \eta_{ip\sigma}\right)^T$
The composite Green's function is thus defined as
\begin{align}
    \mathds{G}(t)=-i\Theta(t) \left\langle \{ \mathbf{\Psi}(t ) , \mathbf{\Psi}^\dagger(0) \} \right\rangle,
\end{align}
where $\Theta(t)$ is the Heaviside theta function. Applying equation of motion for the composite Greens function and assuming that the current $\mathbf{J}$ remains proportional the defined operators basis such that, ${\mathbf{J}(t)=\left[ \mathcal{H} , \mathbf{\Psi}  \right]  (t) \approx \mathds{E} \Psi(t)}$ we obtain the retarded (advanced) Green's function,
\begin{align}
    \mathds{G}^{R/A}(\omega)=\left[ (\omega \pm i0^+) \mathds{1} - \mathds{E} \right]^{-1} \mathds{I},
    \label{Eq:Grealomega}
\end{align}
where $\mathds{1}$ is a $4N\times4N$ unit matrix. Also, we define the M-matrix $\mathds{M}= \left\langle \{ \mathbf{J},\Psi^\dagger \}  \right\rangle$. The I-matrix is given by    ${\mathds{I} = \left\langle \{ \Psi,\Psi^\dagger \}  \right\rangle}$. The energy matrix is then written as
\begin{align}
    \mathds{E}= \mathds{M} \mathds{I}^{-1} \label{Eq:Emat}
\end{align}
For translationally invariant systems we perform a Fourier transform and conduct the calculations in momentum space. Consequently, these matrices become $4\times4$ block diagonal matrices for each $\mathbf{k}$ in the first Brillioun zone in the momentum representation. The Equations of the $\mathds{E}$, $\mathds{I}$ matrices and the self-consistent procedure is presented in Ref.~\cite{Haurie_2024,pangburn2024spontaneous,CDW_COM_AB} and outlined in App.~\ref{Appendix:Details_COM}.

\subsection{Winding number $N_3$}
We can compute the electronic Green function $\mathcal{G}$ from the corresponding relation with the composite Greens function. The topological invariant of a Chern insulator is given by the Chern number $\mathcal{C}$ which is proportional to the Hall conductivity $\sigma_{xy}=\mathcal{C} (e^2/h)$. However, it is more straightforward to calculate an invariant $N_3$, which uses the single particle Greens function
\begin{align}
N_3\big[\mathcal{G}\big]=&\sum_{\alpha \beta \gamma}\dfrac{\epsilon_{\alpha\beta\gamma}}{6}\int_{-\infty}^{+\infty}d\omega \int \dfrac{d^2k}{(2\pi)^2} \label{eq:N3_winding}
 &\\
\nonumber& \times \text{Tr}\big[(\mathcal{G}^{-1}\partial_{k_\alpha}\mathcal{G})(\mathcal{G}^{-1}\partial_{k_\beta}\mathcal{G})(\mathcal{G}^{-1}\partial_{k_\gamma}\mathcal{G})\big]&
\end{align}
where $\epsilon_{\alpha\beta\gamma}$ is the Levi-Civita symbol, and the indices take values $\{0,1,2\}$ with $k_0 = \omega$ and $k_1, k_2$ as components of the crystal momentum. This framework was developed for interacting Green's functions~\cite{gurarie2011single} and is exactly equal to the Chern number in non-interacting systems such that $N_3 = \mathcal{C}$. However, for interacting systems, the relation between $N_3$ and $\mathcal{C}$ does not generally hold. The formula can fail in the presence of Greens function zeros in the gap and also for degenerate ground state~\cite{he2016topological1, he2016topological2,zhao2023failure}. 

In the following analysis, we use $N_3$ to compute the topological invariants associated with the Hubbard excitation bands and the Green function zeros, ensuring that the chemical potential is carefully tuned to accurately capture $N_3$ for the specific bands of interest. Although the value of $N_3$ depends on the chemical potential rather than the filling due to the presence of the Mott gap~\cite{zhao2023failure}, it remains a well-defined mathematical procedure to analyze the topology of each isolated band of zeros. However, a non-trivial value of $N_3$ obtained in this way does not necessarily carry direct physical significance. This approach is valid because $N_3$ is additive, meaning that for two disconnected bands, $b_1$ and $b_2$, the total invariant is simply the sum of the invariants for each band,
$N_3\left[b_1\oplus b_2\right]=N_3\left[b_1\right]+N_3\left[b_2\right].$ 

We adopt a paramagnetic approximation, under which the Hubbard bands are spin-degenerate. Furthermore, in the absence of spin-orbit coupling, the spin sectors remain fully decoupled. As a result, the actual topological invariants of the system are twice the values computed within each spin sector. In the following, we denote the topological invariants for each spin sector as $\mathcal{C}_\sigma$ and $N_{3,\sigma}$, with the understanding that the total topological invariants of the full system are given by $\mathcal{C} = 2\mathcal{C}_\sigma$ and $N_3 = 2N_{3,\sigma}$.

The many-body Chern number can be determined exactly by analyzing the boundary spectrum. Additionally, we evaluate it using the St\v{r}eda formula, as detailed in Appendix~\ref{App:Streda}, which allows for a direct computation based on the particle density response to a small magnetic field~\cite{streda1982theory,hofstadter1976energy}
\begin{align}
&\sigma_{xy}=\phi_0\dfrac{\partial n}{\partial B}|_{\mu,T=0}&
\end{align}
with $\phi_0$ the magnetic quantum of flux.

\subsection{Inhomogeneous systems}
We numerically solve the self-consistent equations detailed in Appendix~(\ref{App:Real_SpacE_Matrix}) for inhomogeneous systems within a real-space framework. Specifically, we analyze systems with open boundary conditions, using translational invariance along the periodic direction. Additionally, we investigate perturbations in clean systems by introducing local defects, such as static $\pi$-fluxes. These defects are incorporated by adding a Peierls phase to all hopping terms that cross a line connecting two magnetic $\pi$-fluxes~\cite{assaad2012topological,jurivcic2012universal,slager2019translational,schindler2022topological}. When the two $\pi$-fluxes are sufficiently far apart, their interactions can be neglected, allowing us to treat them as isolated $\pi$-fluxes. Such local defects acts as local probe for topology for non-interacting systems. 

\section{Results}
\label{sec:Res}
Our calculations are on a square lattice with a linear dimension $L=300$ for the translation-invariant system.The interaction strength on the $p$-orbitals is set to $U_p = 13t$. However, we have ensured the reliability of our qualitative conclusions across various repulsion strengths ranging from $U_p=8t$ to $U_p=20t$. The system is studied at half-filling. However, to mitigate convergence issues of fixing the chemical potential in a large Mott gap, we introduce a small hole doping level such that $n = 1.99$. After obtaining self-consistent parameters, we fix the chemical potential so that the lower subbands are fully occupied, consistent with a half-filled system. The topological features of the Hubbard bands and the zeros of the Green's function cannot be affected by small deformations of the self-consistent parameters. Furthermore, we have benchmarked the electronic Green's function obtained from the COM with the exact solutions in the Hubbard dimer system in Appendix~\ref{App:Dimer}.

We perform the calculations for smaller system sizes for the translation symmetry broken systems. Specifically, for systems with \( \pi \)-flux defects See App.~(\ref{App:Pi}) we use a $50 \times 50$ lattice, while for open boundary conditions, we simulate a \( 120 \times 120 \) system, exploiting translational invariance in one direction. The real (or hybrid) space calculations are initialized using the parameters from the translationally invariant system. In the main text, we present single-shot calculation results highlighting the bulk-boundary correspondence for poles and zeros. We also performed fully self-consistent calculations, which yielded similar bulk-boundary correspondence, with inhomogeneous self-consistent parameters.

\subsection{Topological Mott Band Insulator}
First, consider the non-interacting Chern insulator case $( U_s = U_p = 0)$, as defined in Eq.~(\ref{eq:Hamiltonian}). In this limit, the system is topologically non-trivial, with a non-zero Chern number for $-4t < M < 4t$, and gap closures occurring at $ M = \{-4t, 0, 4t\}$~\cite{bernevig2006quantum}. In contrast, when interorbital hopping is absent, and electron-electron repulsions are strong, each orbital splits into lower and upper Hubbard subbands, separated by the respective interaction strengths $U$. Previous studies have demonstrated that simple intraorbital hopping can induce an selective Mott phase~\cite{pangburn2024spontaneous}. Next, we investigate whether the dispersion of charge excitations in a Mott insulator can acquire a non-zero winding  due to the Hubbard subband inversion which we term as Topological Mott band Insulator (TMBI).
\subsubsection{Single-electron excitation spectrum}
\label{sec:Bands}
\begin{figure}[h!]
\includegraphics[width=8.5cm,keepaspectratio]{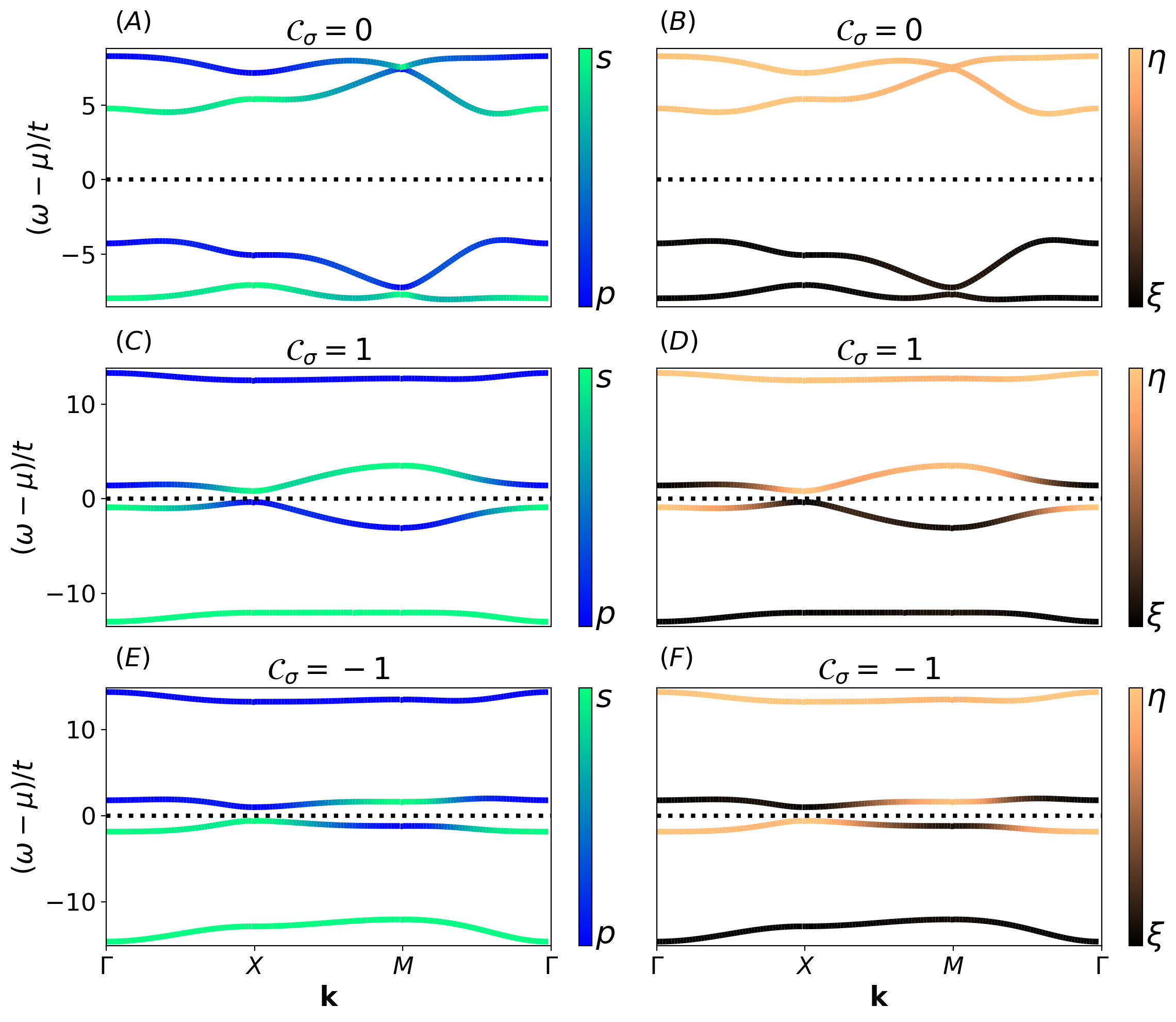}
\caption[0.5\textwidth]{Hubbard subbands along the path \( \Gamma \to X \to M \to \Gamma \) for \( U_s = 12.2t \): (A) \( M = -0.28t \), (B) \( M = 4.36t \), and (C) \( M = 6.08t \). In the left column, colors represent the orbital character of each band, while in the right column, colors denote the holon-doublon excitation character. Bands below the black dotted line indicate filled states. In the trivial topological phase (A), the two occupied bands are primarily holon-like, with opposite orbital characters at high-symmetry points. In the non-trivial topological phases (B/C), the occupied band near the gap exhibits a mixed holon-doublon character, which affects its orbital character and winding number.}
\label{fig:BandHybridization}
\end{figure}

Generally, the winding number $N_{3,\sigma}$ is calculated from the electronic Green's function using Eq.~(\ref{eq:N3_winding}). However, an alternative approach allows us to intuitively infer the topological nature of the bands. For a non-interacting Hamiltonian that is symmetric under inversion \( \mathcal{I} \) and fourfold rotation \( \mathcal{C}_4 \), the parity of the Chern number can be obtained by evaluating the product of the inversion eigenvalues \( \lambda^{\mathcal{I}}_\mathbf{k} \) of the occupied bands at the high-symmetry points \( \Gamma \) and \( M \) in the first Brillouin zone (FBZ)~\cite{fu2007topological, alexandradinata2014wilson}. This is given by
\begin{align}
(-1)^{\mathcal{C}_\sigma \bmod 2} = \prod_{\mathbf{k} \in \{\Gamma, M\}_{\text{occ}}} \lambda_{\mathbf{k}}^{\mathcal{I}},
\label{eq:Fu_Kane}
\end{align}
where the product is taken over all occupied states at $\Gamma,M$ points. The band becomes topological if the product $\lambda_{\mathbf{k}}^{\mathcal{I}}$ becomes negative. The criterion remains applicable for the interacting system, provided that the excitations exhibit a sufficiently long lifetime. For paramagnetic solutions with only holon-doublon excitations, such excitations are infinitely long-lived~\cite{roth1969electron}.

In Fig.~(\ref{fig:BandHybridization}), we plot the quasiparticle excitation bands of the interacting Chern-Hubbard model, displaying their orbital characters (\( s \) and \( p \)) in the left panels and their quasiparticle excitation characters (\( \xi \) and \( \eta \)) in the right panels. As expected for a half-filled system, two of the four excitation subbands lie entirely below the chemical potential, indicated by the dotted lines in Fig.~(\ref{fig:BandHybridization}). For \( U_\alpha \gg M \), the lower Hubbard subbands of both orbitals fall below the Fermi level, as shown in Fig.~(\ref{fig:BandHybridization} A). Due to interorbital hopping, the holon bands can undergo orbital hybridization, giving rise to mixed orbital character at high-symmetry \( \mathbf{k} \)-points. Notably, the product of inversion eigenvalues remains positive, indicating that the occupied bands are topologically trivial. This trivial phase allows for a smooth deformation in which the lowest subband acquires pure \( s \)-character, while the subband just below the chemical potential becomes purely \( p \)-character.

Next, we increase the mass term \( M \) in Fig.~(\ref{fig:BandHybridization} C), keeping all other parameters the same as in Fig.~(\ref{fig:BandHybridization} A). Under this change, the energy gap between the upper and lower Hubbard subbands closes and then reopens at the \( \Gamma \)-points of the FBZ. This transition leads to an inversion of the parity eigenvalue, switching from \( s \)-character (\( \lambda^{\mathcal{I}} = 1 \)) to \( p \)-character (\( \lambda^{\mathcal{I}} = -1 \)), as illustrated in Fig.~(\ref{fig:BandHybridization}). Consequently, the product of the parity eigenvalues \( \lambda^{\mathcal{I}}_{\mathbf{k}} \) for the occupied subbands at the \( \Gamma \) and \( M \) points becomes negative. According to Eq.~(\ref{eq:Fu_Kane}), this indicates that the charge excitation spectrum of the Mott insulator acquires a non-trivial, odd Chern number.

In this process, the lower Hubbard subbands involved in the gap closing acquire a mixed holon-doublon character.  Specifically, the holon of the \( p \)-orbital hybridizes with the doublon excitation of the \( s \)-orbital, as illustrated in Fig.~(\ref{fig:BandHybridization} D), enabling an exchange of inversion eigenvalues between the orbitals.  This hybridization provides the only mechanism for realizing a non-trivial topological invariant from charge excitations in half-filled Hubbard systems.  Without this interorbital mixing between composite excitation subbands, each would retain a trivial topological invariant \( N_{3,\sigma} \) at half-filling.  We refer to the resulting state, characterized by non-trivial topological Hubbard subbands, as a Topological Mott Band Insulator.

Upon further increasing \( M \), the gap closes and reopens at the \( X \)-point, resulting in another exchange of inversion eigenvalues, as shown in  Fig.~(\ref{fig:BandHybridization} C). However, the product of the inversion eigenvalues for the occupied subbands remains negative, hence indicating the non-zero winding of the occupied band, though now with an opposite winding number. Finally, as \( M \) increases further, a gap closing and reopening occurs at the \( M \)-point. This transition results in a trivial filled band insulator, where the \( s \)-orbital is fully occupied and the \( p \)-orbital remains empty.

While we presented an argument for detecting non-trivial band topology using inversion \( \mathcal{I} \) and \( \mathcal{C}_4 \) rotation symmetry, these symmetries are not strictly necessary for the existence of Chern insulators. Within the ten-fold classification scheme, Chern insulators belong to class A of topological insulators~\cite{chiu2016classification,AltlandZirnbauer} akin quantum hall systems and are therefore more general.

\subsubsection{Bulk boundary correspondence of poles}
Since $N_{3,\sigma}$ is a single-particle topological invariant quantized without symmetry protection, it cannot change without the single-particle gap closing. Consequently, the spectrum must be gapless under open boundary conditions if the system possesses a non-zero winding of the occupied bands or at a junction between regions with differing winding numbers. Here, we demonstrate that the bulk-boundary correspondence remains valid in the presence of strong correlations for the TMBI phase.
\begin{figure}[h!]
\centering
\includegraphics[width=8.5cm]{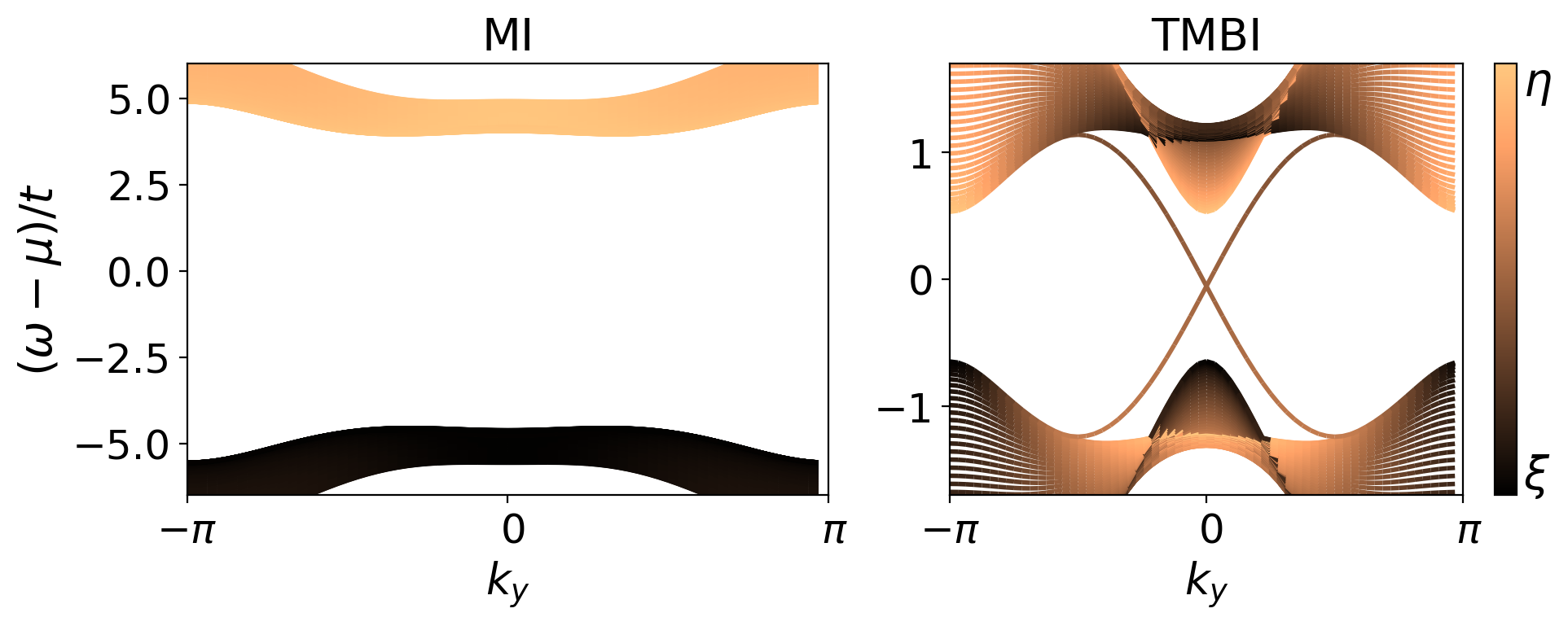}
\caption{Low energy charge excitations of two-band Chern-Hubbard model with open boundary conditions (OBC) with the colors representing the holon-doublon character. (A) Mott Insulator (MI) phase $U_s=12.2 t$ $M=-0.28 t$. (B) Topological Mott Band Insulator phase $U_s=12.2t$ $M=4.36 t$. For full set of COM parameters see Appendix \ref{App:Params}.}
\label{fig:fig2}
\end{figure}

In Fig.~(\ref{fig:fig2}.c/d) we show the electronic excitations eigenvalues with open boundary conditions. As expected from bulk-boundary correspondence, the spectrum is gapped in the MI phase (c), whereas a gapless mode appears in the TMBI phase (d). Since the band inversion in the TMBI phase arises from holon-doublon mixing from the two orbitals, the edge states in this phase also exhibit a mixed holon-doublon character along with a mixed orbital character. Since there are electronic states carrying charge inside the gap this imply a non-zero Hall conductivity $\sigma_{xy}$ which is confirmed using Str\v{e}da formula in Appendix~(\ref{App:Streda}). 

\subsection{Green function zeros}
\label{sec:GreenZeros}

The splitting of the electronic weight between the lower and the upper Hubbard bands lead to frequency $\omega$ satisfying $\text{Det}\left[\mathcal{G}(\mathbf{k},\omega)\right]=0$ leading to a band of Green function zeros~\cite{seki2017topological}. In the atomic limit at half-filling, the zero mode is exactly located half-way between the upper and lower Hubbard bands, as can be seen from the electronic Green function
\begin{align}
&\mathcal{G}_{at}(\omega)=\dfrac{1/2}{\omega-U/2}+\dfrac{1/2}{\omega+U/2}&
\end{align}
By adding hopping $t$ as a perturbation, the zeros acquire some dispersion. Using an approach based on a moments expansion of the Green’s function~\cite{harris1967single}, Wagner and al.~\cite{wagner2023Mott} showed that the band of zeros can be describe by a non-interacting Hamiltonian $H_0(\mathbf{k})$ which  possess the same form than the original non-interacting Hamiltonian with renormalized hopping parameters by interactions. In composite operator formalism we provide a conjecture of GFZ arising from the coherent superposition of holon-doblon pairs in momentum space in App.~(\ref{Appendix:Zeros_Conjecture}). The Greens function zeros emerge in the electronic picture as the tightly bound pairs (the energy for such zeros lies the Mott gap) of elementary excitations holons and doublons.

At each zeros satisfying $\text{Det}\big[\mathcal{G}(\mathbf{k},\omega)\big]=0$, there is an associated zero modes $\ket{\phi_0}$ such that $\mathcal{G}(\mathbf{k},\omega)\ket{\phi_0} = 0$ where the zero mode can be calculated up to an arbitrary $U(1)$ phase. We numerically checked that the gauge invariant quantities computed with $\ket{\phi_0}$ agrees with the same gauge invariant quantities computed from the eigenvector of an effective zeros-Hamiltonian $H_{0}(\mathbf{k})$, defined in  App.~(\ref{Appendix:Zeros_Conjecture}). The eigenvalues of $H_{0}$ is the zeros-dispersion computed from the electronic Greens function. Furthermore, $H_{0}(\mathbf{k})$ has the same form as the non-interacting Hamiltonian but with renormalized tight-binding parameters from interactions in agreement with continued fraction expansion~\cite{wagner2023Mott} and slave-rotor~\cite{wagner2023edge} methods. 

\subsubsection{Dispersion of Green function zeros}
\begin{figure}[h!]
\centering
\includegraphics[width=8.5cm]{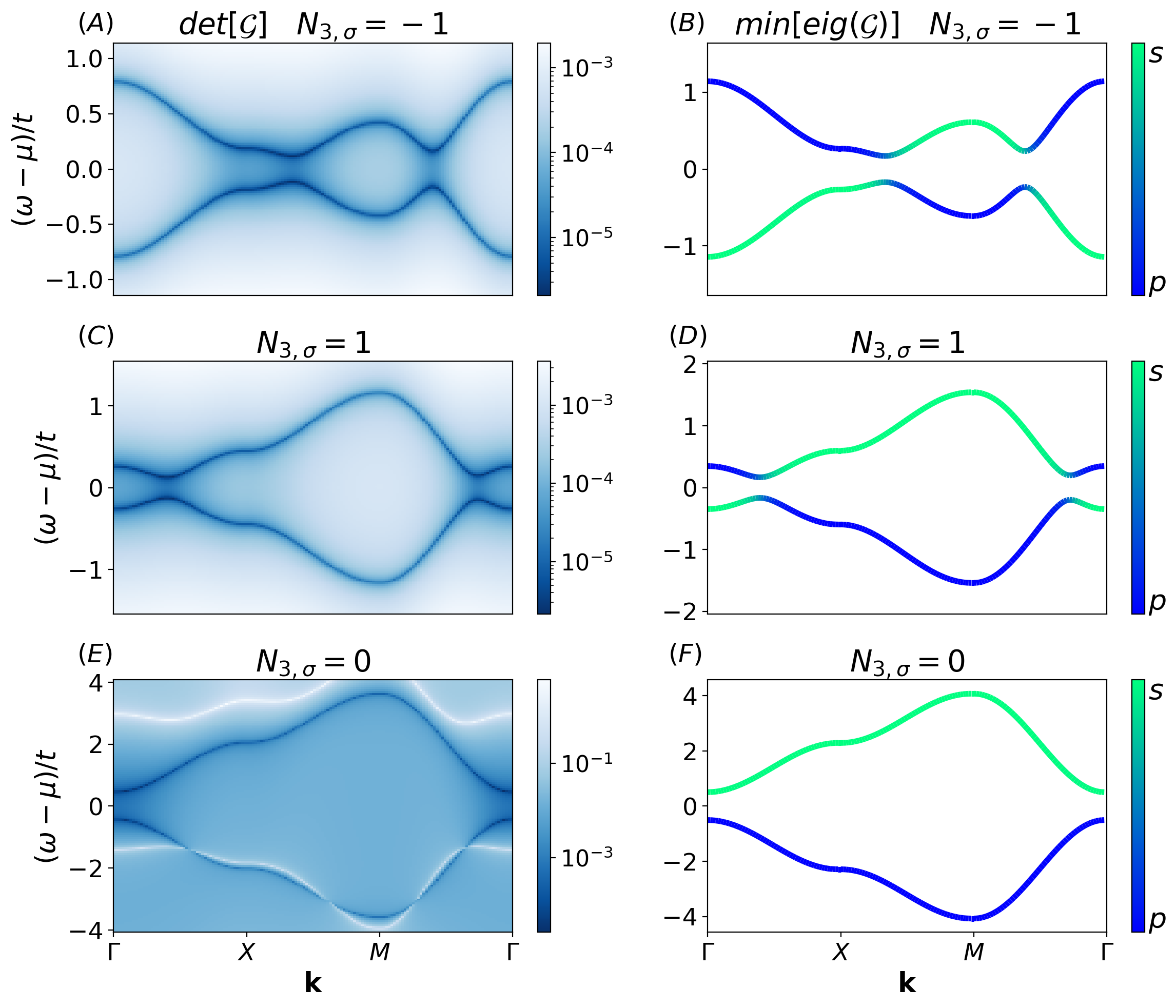}
\caption{Green function zeros are shown for the following cases :  (A/B) $U_s=13.7 t$ $M=-0.48 t$, (C/D) $M=0.53 t$ $U_s=13.7 t$ and (E/F) $M=0.53 t$ $U_s=7.6t$. The left side displays $\text{Det}[\mathcal{G}(\mathbf{k},\omega)]$ along the path $\Gamma X M \Gamma$. The right side shows the orbital character of the electronic Green function zero modes. In the non-zero winding (A-D), the lower zero band exhibits a mixed orbital character, showing band inversion at the $X$ and $M$ points. In contrast, in the trivial phases (E/F), the lower zero band maintains a uniform orbital character.}
\label{fig:fig3}
\end{figure}
In Fig.~(\ref{fig:fig3}), we examine the zeros of the Green's function by computing \( \mathcal{Z}(\mathbf{k}, \omega) = \text{Det}\big[\mathcal{G}(\mathbf{k}, \omega)\big] \) along the high-symmetry path \( \Gamma \rightarrow X \rightarrow M \rightarrow \Gamma \). The manifold of zeros can be identified by locating the minima of \( \mathcal{Z} \) at each crystal momentum and frequency.

In the Chern-Hubbard model, two orbitals with strong on-site repulsion yield two bands of zeros due to interactions. In Fig.~(\ref{fig:fig3}b/d/f), we display the orbital character of these zeros for various parameters by finding the eigenvectors corresponding to the zero eigenvalue of the Green's function. This orbital characterization can also be done by  diagonalizing \( H_{0}(\mathbf{k}) \) (see Appendix.~(\ref{Appendix:Zeros_Conjecture})).

The winding number associated with the lowest zero bands can be computed directly using using \( N_{3,\sigma}\left[\mathcal{G}\right] \) or \( H_{0}(\mathbf{k}) \). Since the Hamiltonian describing the bands of zeros is a local tight-binding model, the upper and lower bands of zeros possess opposite winding numbers, canceling the overall winding number across the two bands. Between Fig.~(\ref{fig:fig3}c) and Fig.~(\ref{fig:fig3}d), the zero spectrum gap closes at the \( X \)-point. Similar to the Green's function poles, this gap closing flips the sign of \( N_{3,\sigma} \) while preserving a non-zero winding number. Fig.~(\ref{fig:fig3}e) shows an example of a topologically trivial zero band with \( N_{3,\sigma} = 0 \), where each band of zeros has a distinct orbital character.

However, it is important to note that the chemical potential for such computations are set within the gap of zeros to perform the computation of the winding number. Generally at zero temperature, the chemical potential can be placed anywhere within the electronic excitation gap without altering any physical quantity. The filling of a specific GFZ band by a particular choice of \( \mu \) remains physically ambiguous but is mathematically allowed. However, when the zero bands acquire a finite winding number due to the positioning of the chemical potential between them, we refer to this phase as the Topological Mott Zero (TMZ).

\subsubsection{Bulk boundary correspondence of Greens function zeros}
\begin{figure}[h!]
\centering
\includegraphics[width=8.5cm]{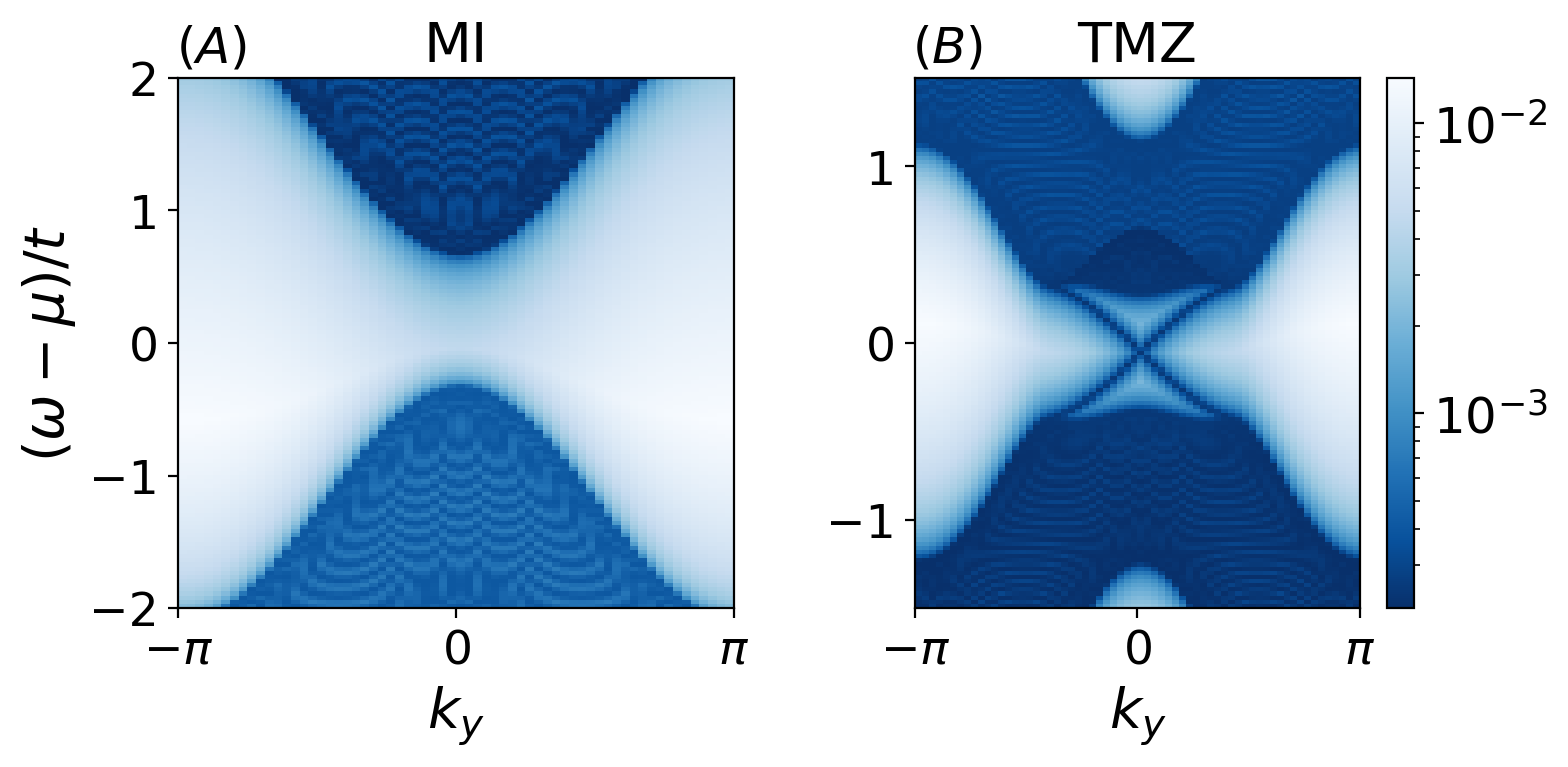}
\caption{The minimum eigenvalue of the single particle electronic Green function $\mathcal{G}$ is shown in the Mott Insulator phase (A) $M=0.53 t$ $U_s=7.6t$  and Topological Mott Zeros phase (B) $M=0.53 t$ $U_s=13.7 t$. $min\left[eig(G)\right]$ is shown with open boundary conditions in function of $\omega$ and the transverse momentum $k_y$. As expected from the bulk-boundary correspondence of zeros, the TMZ phase exhibits gapless edge states with OBC while the trivial case stay gapped under OBC. For full set of COM parameters see Appendix \ref{App:Params}}.
\label{fig:fig4}
\end{figure}

Since the zeros band can carry a non-zero $N_3$ invariant, there exists a `bulk-boundary' correspondence for GFZ, similar to the well-known correspondence for poles. In Fig.~(\ref{fig:fig4} A) and Fig.~(\ref{fig:fig4} B), we show the minimum eigenvalue of $\mathcal{G}(\omega,k_y)$ with open boundary condition in the MI phase and the TMZ phase respectively. As anticipated, the zeros spectrum is gapped in the MI case, while it remains gapless in the TMZ case. Note that GFZ is a charge neutral mode, so it cannot generate a conducting edge. Since the system has gapped excitation energy and the GFZ cannot contribute to the Hall conductivity $\sigma_{xy}$. We have found that the Hall conductivity vanishes for the TMZ using Str\v{e}da formula. Thus, for the TMZ phase with open boundary conditions is insulating both in the bulk and the edge.

\subsection{Phase Diagram}
\begin{figure}[h!]
\centering
\includegraphics[width=8.5cm]{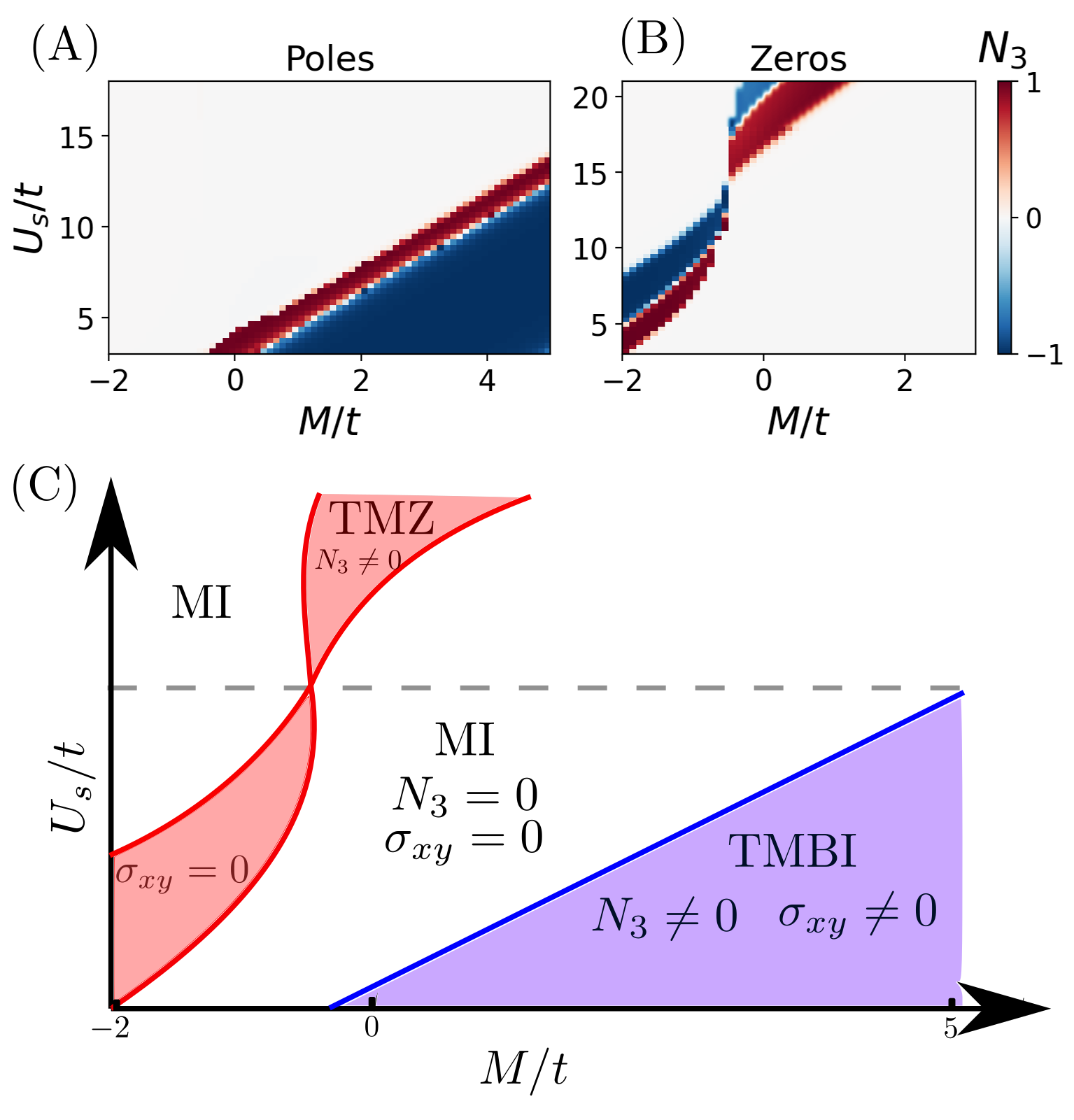}
\caption{(A) Displays the $N_3$ invariant of the electronic bands in function of $U_s$ and $M$. (B) Presents the $N_3$ invariant of the zeros bands in function of $U_s$ and $M$. (C) Summarizes the topological phase diagram of the Chern-Hubbard model in function of $U_s$ and $M$.The dotted gray line indicates $U_s=U_p$.}
\label{fig:fig5}
\end{figure}
Here, we present the phase diagram of winding numbers for the poles and zeros of the two-point Green's functions. This phase diagram is determined by fully solving the self-consistent equations in momentum space and calculating the winding number \( N_3 \) from the electronic Green's function, with \( U_p = 13t \) fixed.

In Fig.~(\ref{fig:fig5} A), we show the winding number for the electronic bands when the system is at half-filling. First we focus where both orbitals have large interactions compared to the crystal field splitting. In these regimes, the gap between the lower and upper Hubbard bands is substantial, preventing subband hybridization. Therefore, finite winding number sets in only at large values $M$. As the interaction strength \( U_s \) increases, the required crystal field potential required for subband hybridization decreases. Finally, increasing the crystal field potential while keeping \( U_s \) fixed closes the energy gap, rendering the winding number ill-defined within a narrow range of \( M \). Upon further increase in \( M \), a topological transition occurs to a phase with a winding number of opposite parity. As expected for lower interactions, the value at which the model attains topological non-trivial character approaches the non-interacting limit. However, the formation of doublons and holons requires the presence of strong interactions compared to hopping, and this regime remains inaccessible in a strong coupling approach.

Fig.~(\ref{fig:fig5} B) illustrates the winding number phase diagram for the bands of Green's function zeros, indicating the parameter regime where the bands of zeros exhibit non-zero winding. Notably, the bands of GFZ acquire a non-zero $N_{3,\sigma}$ in a parameter regime that differs entirely from the poles. As the gap in the Green's function zeros closes, the winding number becomes ill-defined and reverses sign when the GFZ gap reopens. Increasing the crystal field splitting renders the zeros trivial. For $U_s=U_p$ the regime for which the GFZ has a finite winding almost vanishes. The $N_{3,\sigma}$ of zeros below and above this interaction is approximately symmetric.

Finally, Fig.~(\ref{fig:fig5} C) summarizes a schematic phase diagram for a two-orbital Mott insulator with complex interorbital hopping. The white regions represent a trivial Mott insulating phase, while the violet regions denote regimes where the single particle excitation bands have non-zero winding. The system is in the quantum anamolous Hall regime and supports a gapless conducting mode at the edge via bulk-boundary correspondence.

Moreover, the zero bands can support non-zero winding numbers in a distinct parameter regime. In the Mott insulating phase, we show that such bands of GFZ yield gapless charge-neutral modes at a boundary. In contrast, the electronic bands remain topologically trivial, ensuring the Hall conductivity vanishes due to the absence of boundary charge mode. 

With interactions, the correspondence between the single-particle Green function winding number $N_{3,\sigma}$ and the Hall conductivity $\sigma_{xy}$ no longer holds, as GFZ do not contribute to $\sigma_{xy}$~\cite{zhao2023failure}. Given that the zeros and poles in this model attain non-zero winding numbers in entirely separate regimes, we further examine the interaction between edge modes from the electronic bands and the gapless modes of Green's function zeros, motivated by Refs.~\cite{wagner2023Mott,wagner2023edge}. 
\subsection{Junction between TMBI and TMZ phases}
\begin{figure}[h!]
\includegraphics[width=8cm]{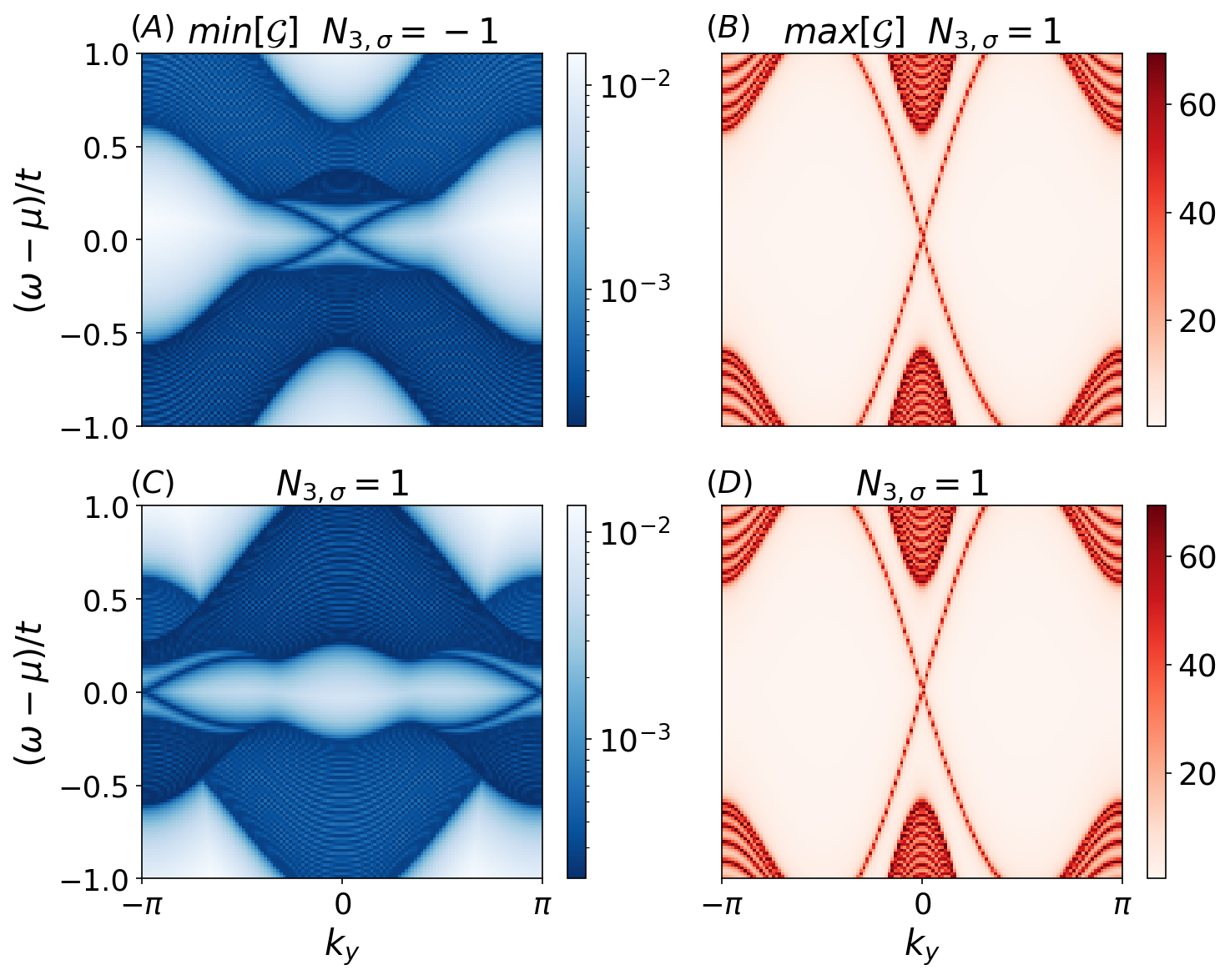}
\caption{The minimum (left column) and maximum (right column) eigenvalue of the single particle Green function $\mathcal{G}$ are shown in function of $\omega$ and $k_y$. Both (A/B) and (C/D) corresponds to the same single-particle Green function $\mathcal{G}$. In both cases, we select a representative set of self-consistent parameters for a system divided into two halves. One half is in the TMBI phase with $N_{3,\sigma}=1$, while the other half is in the TMZ phase, with $N_{3,\sigma}=-1$ in the upper row (A) and $N_{3,\sigma}=1$ in the lower row (C). For full set of parameters see Appendix \ref{App:Params}.}
\label{fig:fig6}
\end{figure}

We design a hypothetical tunnel junction such that the parameters are chosen on one side to model the TMZ phase, and the another side models the TMBI phase with the same or opposite winding number. In the Fig.~(\ref{fig:fig6}) top panels the terminal on the left side (say) is in TMZ phase with $N_{3,\sigma}=-1$  and the terminal on the right TMBI with $N_{3,\sigma}=1$. The spectrum shows a gapless zero mode Fig.~(\ref{fig:fig6}A) and a gapless edge state in Fig.~(\ref{fig:fig6}B). Similarly when the winding number for both TMZ and TMBI phase are the same, i.e. $N_{3,\sigma}=1$, we still observe a gapless edge state and gaples zero mode (See Fig.~(\ref{fig:fig7}C) and Fig.(\ref{fig:fig7}D)). The parameters corresponding to the different phases are listed in App.~\ref{App:Params}.

We showed in the previous section that in the TMZ phase, where there is a nonzero winding number, the Hall conductivity \( \sigma_{xy} \) still vanishes as the system has a single-particle excitation gap. In contrast, a Topological Mott Band Insulator phase exhibits a nonzero \( \sigma_{xy} \) due to the conducting edge states. Consequently, bulk-boundary correspondence requires a gapless edge state at any boundary where the Hall conductivity changes. In any formalism restricted to the single-particle Green’s function, the \( U(1) \) charge is carried by electrons. As a result, edge poles cannot annihilate with edge zeros, as such a possibility would violate the fundamental law of charge conservation. Furthermore, the results reiterate that the correspondence between the $N_{3,\sigma}$ and the Hall conductivity breaks down entirely for interacting systems. To accurately gauge the topological invariant $\sigma_{xy}$ can be calculated by integrating the Berry curvature over the twisted boundary condition parameters~\cite{sinha2024computing} or using the standard Kubo formalism. In App.~\ref{App:Streda} we perform the calculation of $\sigma_{xy}$ using Str\v{e}da formula. In App.~\ref{App:Junction}, we verify that the bulk-boundary correspondence also holds at the junction between a TMZ and MI phase, as well as between a TMBI and MI phase.

\section{Conclusions and Discussions\label{sec:Discussions}}

\subsection{Relationship of TMBI to non-interacting band topology}
Electronic excitations in the strong repulsion regime split into lower and upper Hubbard bands dominated by holons and doublons respectively. Here, we demonstrate that two strongly correlated orbitals can undergo topological band inversion facilitated by the crystal field splitting potential. This mechanism enables mixing between the upper and lower Hubbard bands of separate orbitals, driving a transition from a trivial Mott insulator to a quantum anomalous Hall insulator. Our self-consistent calculations provide evidence for interaction-induced topological phases on a square lattice Chern insulator across an extensive parameter regime. These phases bear similarity to those in the Kane-Mele-Hubbard model~\cite{mai2024topological}. The mechanism resembles the topological band inversion between lower Hubbard and charge-transfer bands discussed in Ref.~\cite{devakul2022quantum} rather than between the upper and lower Hubbard bands of the different orbitals studied here.

Within the composite operator formalism, the lower and upper Hubbard bands correspond primarily to holons and doublons, which behave as quasi-free quasiparticles. This allows the study of topological invariants to parallel non-interacting band topology~\cite{bradlyn2017topological}, albeit in holons and doublons, where interactions renormalize their effective parameters. The strong correlations manifest primarily in the electronic framework.

This simplified picture is valid when the proposed holon-doublon excitations have sufficiently long lifetimes, as in the paramagnetic Mott insulator phase, where these excitations are infinitely long-lived. While we anticipate that the results extend to cases with weak antiferromagnetic ordering at half-filling, a detailed analysis of such scenarios remains open for future investigation.

\subsection{Topological Green function zeros of Mott insulator}
The electronic Green's function remains strongly correlated, as the electronic spectral weight is split between the upper and lower Hubbard bands with mixing~\cite{hirsch1992superconductors}. By analyzing holon-doublon excitations, we reproduce these Green's function zeros with a conjectured effective Hamiltonian of zeros. As demonstrated in Appendix~(\ref{Appendix:Zeros_Conjecture}), the COM formalism provides an interpretation of Green's function zeros as tightly pairs of holon and doublon elemetary excitations. These zeros emerge in the electronic framework after integrating out elementary $\xi$ and $\eta$ excitations, supporting the view that electrons in strongly correlated Mott insulators behave as composite particles~\cite{Weinberg}.

Furthermore, the GFZ bands can acquire topological character, identified by a nonzero winding number $N_{3,\sigma}$. The parameter regime where GFZ bands exhibit nonzero winding numbers differs significantly from the non-interacting Chern model model, as the strong correlations renormalize the band structure. A system with a nonzero winding number necessarily features gapless GFZ modes due to bulk-boundary correspondence. However, single-particle electronic excitations remain gapped in the Topological Mott Zero phase, rendering the system insulating. As a result, single-particle electronic quasiparticles cannot carry a Hall current at the boundary. This implies that the winding number $N_{3,\sigma}$, calculated from the single-particle Green's function, does not correspond to the physical topological invariant, namely the Chern number derived from the Hall conductivity.

\subsection{Interaction between the poles and zeros}
We also investigated a proposed junction geometry with different winding numbers between the TMBI and TMZ phases. Let us first consider the case where the TMBI and TMZ phases exhibit different winding numbers. Since the winding number transitions from one region to another, such a change necessitates a closure of the single-particle gap. Consequently, we observe both a boundary edge state and a boundary edge zero mode. Similarly, when the TMBI and TMZ phases share the same winding number, gapless edge states and gapless GFZ modes are still observed at the boundary. Despite these gapless zero modes, the TMZ phase remains an trivial insulator with no Hall conductivity, while the TMBI phase exhibits a finite Hall conductivity. Within the single-particle Green's function framework, electron-like quasiparticles carry charge; thus, edge poles cannot annihilate edge zeros. The only scenarios in which the edge states associated with poles can become gapped are either through the closure of the many-body gap~\cite{mai2023topological} or due to non-trivial topology arising in multi-particle Green's functions, both of which extend beyond single-particle descriptions.

Related studies~\cite{wagner2023edge, wagner2023Mott} have examined the interface between a non-interacting topological insulator and a TMZ phase. In Ref.~\cite{wagner2023Mott}, the SSH chain cancellation of the pole and zero edge states demonstrated via exact diagonalization. Meanwhile, in Ref.~\cite{wagner2023edge}, employing the slave-rotor formalism on the Kane-Mele model, it was shown that spinon edge states become gapped at the boundary between a topological insulator and a TMZ phase. In contrast, rotor edge states remain gapless, facilitating the necessary current between regions with $\sigma_{xy} = 0$ and $\sigma_{xy} \neq 0$.

\subsection{Extension to many-body topology}
In this work, we limit our study to the single-particle Green's function. However, a many-body description is needed to capture topological properties in strongly correlated systems fully. The Hubbard operator formalism provides a framework for extending topological aspects beyond single-particle physics.
 
One potential extension involves incorporating interactions between the fermionic and bosonic sectors, as demonstrated in studies of the $t$-$J$ model~\cite{hassan2024strong}, without expanding the fermionic composite basis. Another approach is to include inter-site composite fermionic excitations within the composite operator basis, as explored in previous work on the single-band Hubbard model~\cite{odashima2005high}. In this framework, electronic excitations encompass atomic-limit excitations and inter-site operators induced by hopping processes. Such extensions remain a challenge for future studies in strongly correlated systems beyond the single particle Greens function.





\section{Acknowledgement}
The authors thank Karyn le Hur, C. Bena for useful discussions. The authors thank B.J. Wieder for suggestions regarding the draft. The authors acknowledges funding from CEPIFRA (Grant No. 6704-3). The calculations are performed on the IPhT Kanta cluster. 
%

\appendix

\begin{onecolumngrid}

\section{Green function zeros in composite operator method \label{Appendix:Green_Zeros}}
\subsection{Single-orbital Hubbard Model}
For single band Hubbard model the dispersion of Green function zeros at a filling $n=1$ can be readily obtained from Ref.~{\cite{Haurie_2024}}. At half-filled case, $e=0$ and hence the M-matrix becomes $2\times2$ matrix and is given by
\begin{align}
\mathds{M}(\mathbf{k})=\begin{pmatrix}
-\mu/2-\gamma(\mathbf{k}) p & -\gamma(\mathbf{k})(1/2-p) \\
-\gamma(\mathbf{k})(1/2-p) & -(\mu-U)/2-\gamma(\mathbf{k}) p \\
\end{pmatrix}
\end{align}
with diagonal $\mathds{I}^{-1}=\left[2,2 \right]$. The composite Greens function can then be constructed. Using that $\mathcal{G}(\mathbf{k},\omega)=\sum\limits_{\alpha,\beta=1}^2\mathds{G}_{\alpha\beta}(\mathbf{k},\omega)$, we can then solve $\forall \mathbf{k}$ $\det\left[\mathcal{G}\left(\mathbf{k},\omega\right)\right]=0$. For each $\mathbf{k}$, there is a single solution $\omega_0(\mathbf{k})$ given by the following expression
\begin{align}
\omega_0(\mathbf{k})&=\left(M_{11}(\mathbf{k})+M_{22}(\mathbf{k})-2 M_{12}(\mathbf{k})\right) \label{Eq:ZerosSingle}&\\
 \omega_0(\mathbf{k})&=2t(1-4p)(\cos(k_x)+\cos(k_y)) +U/2-\mu
\end{align}
The existence and uniqueness of the solution at each $\mathbf{k}$ arise from the general properties of $\text{Re}\left[\mathcal{G}(\mathbf{k},\omega)\right]$, provided the electronic weight is distributed between the upper and lower Hubbard bands. It is a continuous, decreasing function with respect to $\omega$, diverging to $\pm \infty$ as it approaches the lower and upper Hubbard bands. Additionally,the single particle excitations vanishes, hence $\text{Im}\left[\mathcal{G}(\mathbf{k},\omega)\right]=0$. 

\subsection{Two-orbital system}
The dispersion of the zeros are also solvable in the case of the Chern-Hubbard model where the $E$-matrix is a $4\times 4$ matrix for each $\mathbf{k}$. Similar argument than the previous section leads to two solutions for $\det\big[\mathcal{G}(\mathbf{k},\omega)\big]=0$ at each $\mathbf{k}$,

\begin{align}
\omega^\pm_0(\mathbf{k})= &\frac{1}{4} \Bigg(\pm\Big[\left(E_{11}n_s-E_{12}n_s-E_{21}(2-n_s)+E_{22}(2-n_s)-E_{33}n_p+E_{34}n_p+E_{43}(2-n_p)-E_{44}(2-n_p)\right)^2& \nonumber \\
&+4
   (E_{13}n_s-E_{14}n_s-E_{23}(2-n_s)+E_{24}(2-n_s))
   (E_{42}(2-n_p)+E_{31}n_p-E_{32}n_p-E_{41}(2-n_p))\Big]^{1/2}& \nonumber \\
&+E_{11}n_s-E_{12}n_s-E_{21}(2-n_s)+E_{22}(2-n_s)+E_{33}n_p-E_{34}n_p-E_{43}(2-n_p)+E_{44}(2-n_p)\Bigg).
\label{Eq:ZerosTwo}
\end{align}
The equation can be expanded in terms of self-consistent parameters $n_\alpha,e_{\alpha\beta},p_{\alpha\beta}$. 
    
\subsection{Conjecture of a Hamiltonian describing Green function zeroes\label{Appendix:Zeros_Conjecture}}
\subsubsection{Single orbital system}
Let us assume we can prepare the system in a state which is an equal superposition state of the doublons and holons for every crystal momentum. The state is given by
\begin{align}
\ket{\psi_{0}(\mathbf{k})}=\dfrac{1}{\sqrt{2}}\left(\ket{\xi(\mathbf{k})}-\ket{\eta(\mathbf{k})}\right)
\end{align}
The density matrix for such states is defined as $\hat{\rho}_{0}(\mathbf{k})=\ket{\psi_{0}(\mathbf{k})}\bra{\psi_{0}(\mathbf{k})}$. 

Conjecture: The Hamiltonian of Green's function zero is given by the following relation
\begin{align}
    H_0(\mathbf{k})=\text{Tr}\big[\mathds{E}(\mathbf{k})\hat{\rho}_0(\mathbf{k})\big],
\end{align}
where $\text{Tr}$ denotes the trace of the matrix which gives just a scalar number at each $\mathbf{k}$. One can check that this is exactly equal to the dispersion described in Eq.~(\ref{Eq:ZerosSingle}). We comment that since the state is a linear combination of doublons and holons they carry no charge. Further more since states are pure the Von Neumann entropy of such states vanishes.
\subsubsection{Two orbital system}
Similarly one can extend the argument for the two-orbital system. Note that at half-filling due to the crystal field splitting potential the orbital densities can vary even when the total electron density is half-filled. Thus, the state corresponds to the coherent superposition of collective excitations of holon and doublon becomes,
\begin{align}
\ket{\psi_{0}(\mathbf{k})}=\dfrac{1}{2}\left(\sqrt{n_s}\ket{\xi_s(\mathbf{k})}-\sqrt{2-n_s}\ket{\eta_s(\mathbf{k})}\right) \otimes \left(\sqrt{n_p}\ket{\xi_p(\mathbf{k})}-\sqrt{2-n_p}\ket{\eta_p(\mathbf{k})}\right)
\end{align}
We can define the $4\times4$ density matrix $\hat{\rho}_{0}(\mathbf{k})=\ket{\psi_{0}(\mathbf{k})}\bra{\psi_{0}(\mathbf{k})}$. 

Conjecture : The dispersion of the electronic Green's function zeros are given by the eigenvalues of the following Hamiltonian
\begin{align}
H_{0}(\mathbf{k})=\text{Tr}_{\{\eta,\xi\}}\left[\mathds{E}(\mathbf{k})\hat{\rho}_{0}(\mathbf{k})\right]
\end{align}
Here we partially trace over the $\{ \eta,\xi \}$ Hilbert space and hence obtain a $2 \times 2$ matrix $H_0$ for each crystal momentum. The eigenvalues of the $H_0$ reproduces the band of Green's function zeros in Eq.~(\ref{Eq:ZerosTwo}).

At each zero satisfying $\text{Det}\big[\mathcal{G}(\mathbf{k},\omega)\big]=0$, there exists an associated zero mode $\ket{\phi_0}$ such that $\mathcal{G}(\mathbf{k},\omega)\ket{\phi_0} = \ket{0}$. The zero modes $\ket{\phi_0}$ can be determined up to an arbitrary $U(1)$ phase. We numerically confirmed that gauge-invariant quantities computed using $\ket{\phi_0}$ match precisely with those derived from the eigenvectors of the zeros Hamiltonian $H_0(\mathbf{k})$. Furthermore, by analyzing the eigenvectors of $H_0(\mathbf{k})$, we verified that the winding number of the Green's function zeros, as calculated using the $N_3$-formula from $\mathcal{G}$, is consistent with the topological characterization of GFZ obtained from $H_0(\mathbf{k})$.
This perspective provides an intriguing interpretation, identifying Green's function zeros as representing tighly-bound pairs of the holon-doublon states~\cite{prelovvsek2015holon,roy2024signatures}. Note that the elementary excitations holons and doublons have no zeros in the spectrum . However, upon integrating out the such quasiparticles the electronic Green's function obtains zeros. Furthermore, the enerfy of the zerors remains within the energy gap. Our findings suggest that the zeros of the Green's function correspond to tightly bound holon-doublon pairs. For \( M \gg t \), one of the bands becomes more occupied than the other. The superposition state acquires different weights for doublons and holons. This observation hints that the non-zero winding number for the GFZ regime appears around \( M \approx 0 \).

A non-trivial winding number associated with the Green's function zeros would then obstruct exponentially Wannierizing the holon-doublon pairs. Recent proposals~\cite{blesio2018topological,vzitko2021iron,fabrizio2022emergent,skolimowski2022luttinger,blason2023unified,fabrizio2023spin,blason2024luttinger,pasqua2024exciton} interpret the Luttinger surface~\cite{dzyaloshinskii2003some} as the Fermi surface of exotic quasiparticles that carry no charge, thereby not contributing to the optical conductivity. This interpretation aligns with the topological invariant associated with Green's function zeros, which is not directly linked to quantized transport coefficients~\cite{zhao2023failure}. However, within this framework, the zeros originate from pure quantum states and thus cannot carry von Neumann entropy.

\section{Details of composite operator formalism\label{Appendix:Details_COM}}
\subsection{Non-uniform systems\label{App:Real_SpacE_Matrix}}
The study of local probes requires a real-space formulation of the composite operator formalism. We first choose the following $4N$-compoment basis for the composite operator
\begin{align}
    \mathbf{\Psi}_i=\left(\hdots ,\xi_{is\sigma}, \eta_{is\sigma}, \xi_{ip\sigma}, \eta_{ip\sigma}, \hdots \right)^T
\end{align}
For completeness, the non-zero elements of diagonal $I$-matrix in the same basis is given by
\begin{align}
\mathds{I}=\text{diag}\left[\hdots 1-n_s(i)/2,n_s(i)/2,1-n_p(i)/2,n_p(i)/2 \hdots\right].
\end{align}

Thus, the $M$-matrix can be expressed in real space. The intra-orbital matrix elements read as follows 
\begin{align}
    &\mathds{M}_{i,j} = -\delta_{ij} \left[\mu \left(1-\frac{n_s(i)}{2}\right) + e_{s}(i) - e_{sp}(i)\right] 
    - t \left(1-\frac{n_s(i) + n_s(j)}{2} + p_{ss}(i,j)\right) \label{eq:M11} \\
    &\mathds{M}_{i,j+N} = \delta_{ij}\left[e_s(i) - e_{sp}(i)\right] 
    - t \left(\frac{n_s(j)}{2} - p_{ss}(i,j)\right) \label{eq:M12} \\
    &\mathds{M}_{i+N,j} = \delta_{ij}\left[e_s(i) - e_{sp}(i)\right] 
    - t \left(\frac{n_s(i)}{2} - p_{ss}(i,j)\right) \label{eq:M21} \\
    &\mathds{M}_{i+N,j+N} = -\delta_{ij} \left[\left(\mu - U_s\right) \frac{n_s(i)}{2} + e_s(i) - e_{sp}(i)\right] 
    - t p_{ss}(i,j) \label{eq:M22} \\
    &\mathds{M}_{i+2N,j+2N} = -\delta_{ij} \left[\mu \left(1-\frac{n_p(i)}{2}\right) - e_p(i) + e_{ps}(i)\right] 
    + t \left(1 - \frac{n_p(i) + n_p(j)}{2} + p_{pp}(i,j)\right) \label{eq:M33} \\
    &\mathds{M}_{i+2N,j+3N} = \delta_{ij} \left[-e_p(i) + e_{ps}(i)\right] 
    + t \left(\frac{n_p(j)}{2} - p_{pp}(i,j)\right) \label{eq:M34} \\
    &\mathds{M}_{i+3N,j+2N} = \delta_{ij} \left[-e_p(i) + e_{ps}(i)\right] 
    + t \left(\frac{n_p(i)}{2} - p_{pp}(i,j)\right) \label{eq:M43} \\
    &\mathds{M}_{i+3N,j+3N} = -\delta_{ij} \left[\left(\mu - U_p\right) \frac{n_p(i)}{2} - e_p(i) + e_{ps}(i)\right] 
    + t p_{pp}(i,j) \label{eq:M44}
\end{align}

Let's define $\delta_x$ and $\delta_y$ as vectors connecting one site to its neighboring sites in the $\hat{x}$ and $\hat{y}$ directions, respectively. The inter-orbital matrix elements read for $j=i+ \delta_x$

\begin{align}
    &\mathds{M}_{i,i+\delta_x+2N} = a(i,i+\delta_x) t\left(1-\dfrac{n_s(i)+n_p(i+\delta_x)}{2}+p_{sp}(i,i+\delta_x)\right) \label{eq:M13x} \\
    &\mathds{M}_{i+2N,i+\delta_x} = a(i,i+\delta_x) t\left(1-\dfrac{n_s(i+\delta_x)+n_p(i)}{2}+p_{sp}(i,i+\delta_x)\right) \label{eq:M31x} \\
    &\mathds{M}_{i,i+\delta_x+3N} = a(i,i+\delta_x) t\left(\dfrac{n_p(i+\delta_x)}{2}-p_{sp}(i,i+\delta_x)\right) \label{eq:M14x} \\
    &\mathds{M}_{i+3N,i+\delta_x+N} = a(i,i+\delta_x) t\left(\dfrac{n_p(i)}{2}-p_{sp}(i,i+\delta_x)\right) \label{eq:M41x} \\
    &\mathds{M}_{i+N,i+\delta_x+2N} = a(i,i+\delta_x) t\left(\dfrac{n_s(i)}{2}-p_{sp}(i,i+\delta_x)\right) \label{eq:M23x} \\
    &\mathds{M}_{i+2N,i+\delta_x+N} = a(i,i+\delta_x) t\left(\dfrac{n_s(i+\delta_x)}{2}-p_{sp}(i,i+\delta_x)\right) \label{eq:M32x} \\
    &\mathds{M}_{i+N,i+\delta_x+3N} = a(i,i+\delta_x) tp_{sp}(i,i+\delta_x) \label{eq:M24x} \\
    &\mathds{M}_{i+3N,i+\delta_x+N} = a(i,i+\delta_x) tp_{sp}(i,i+\delta_x) \label{eq:M42x}
\end{align}
with $a(i,i+\hat{x})=i$ and $a(i,i-\hat{x})=-i$.

For $j=i+\delta_y$, the inter-orbital matrix elements read

\begin{align}
    &\mathds{M}_{i,i+\delta_y+2N} = -a(i,i+\delta_y) t\left(1-\dfrac{n_s(i)+n_p(i+\delta_y)}{2}+p_{sp}(i,i+\delta_y)\right) \label{eq:M13y} \\
    &\mathds{M}_{i+2N,i+\delta_y} = a(i,i+\delta_y) t\left(1-\dfrac{n_s(i+\delta_y)+n_p(i)}{2}+p_{sp}(i,i+\delta_y)\right) \label{eq:M31y} \\
    &\mathds{M}_{i,i+\delta_y+3N} = -a(i,i+\delta_y) t\left(\dfrac{n_p(i+\delta_y)}{2}-p_{sp}(i,i+\delta_y)\right) \label{eq:M14y} \\
    &\mathds{M}_{i+3N,i+\delta_y+N} = a(i,i+\delta_y) t\left(\dfrac{n_p(i)}{2}-p_{sp}(i,i+\delta_y)\right) \label{eq:M41y} \\
    &\mathds{M}_{i+N,i+\delta_y+2N} = -a(i,i+\delta_y) t\left(\dfrac{n_s(i)}{2}-p_{sp}(i,i+\delta_y)\right) \label{eq:M23y} \\
    &\mathds{M}_{i+2N,i+\delta_y+N} = a(i,i+\delta_y) t\left(\dfrac{n_s(i+\delta_y)}{2}-p_{sp}(i,i+\delta_y)\right) \label{eq:M32y} \\
    &\mathds{M}_{i+N,i+\delta_y+3N} = -a(i,i+\delta_y) tp_{sp}(i,i+\delta_y) \label{eq:M24y} \\
    &\mathds{M}_{i+3N,i+\delta_y+N} = a(i,i+\delta_y) tp_{sp}(i,i+\delta_y) \label{eq:M42y}
\end{align}

with $a(i,i+\hat{y})=1$ and $a(i,i-\hat{y})=-1$.

\subsection{Correlation function and self-consistency}
The $M$-matrix and the $I$-matrix depends on initally unknown parameters ($e_{\alpha \beta },p_{\alpha \beta},n_\alpha,\mu$) that needs to be determined self consistently. Note that to evaluate the parameter $n_\alpha$ and $e_{\alpha \beta}$ we only need the single particle on-site and nearest neighbor correlation functions. Applying fluctuation dissipation theorem, the correlation function is given in the translation invariant case by
\begin{align}
\mathds{C}=-\frac{1}{2 i \pi} \int d\omega \left[1+\text{tanh}\left(\dfrac{\beta \omega}{2}\right)\right]& \left(\mathds{G}^R(\omega)-\mathds{G}^A(\omega)\right),
\label{eq:FluctuationDissipation}
\end{align}
with $G^{R/A}$ respectively the retarded and advanced Green function. So for two-point correlations for all the sites and composite operators are there in correlation function $\mathds{C}$ whihc is a $4N\times4N$ matrix. We only need the on-site and nearest neighbor correlations. One can read off the on-site correlation function at site $C_0^{ab}(i)$ and $C_1^{ab}(i)$ as the nearest-neighbor correlation function at site $i$ from $\mathds{C}$. Here $a$ and $b$ are integers between $[1,4]$. We can analytically compute the $\omega$ integral in Eq.~(\ref{eq:FluctuationDissipation}) as detailed in the App.~(\ref{Appendix:correlation}) using the spectral decomposition of the $E$-matrix~\cite{avella2011composite}.  

\subsection{Correlation function computation \label{Appendix:correlation}}
We can simplify the  $\omega$ integration in the fluctuation-dissipation theorem of Eq.~(\ref{eq:FluctuationDissipation}) by using a spectral decomposition. First,  the $\mathds{E}$- matrix can be diagonalize
\begin{align}
    \mathds{E}=\mathds{R} \mathds{D} \mathds{R}^{-1}&,
\end{align}
Where $\mathds{R}$ is the matrix of right eigenvectors of $\mathds{E}$ and $\mathds{D}$ is the diagonal eigenvalue matrix. 
Then the composite operators advanced and retarded Green's function becomes
\begin{align}
    \mathds{G}^{R/A}(\omega) &= \mathds{R} \left(\left(\omega\pm i\epsilon\right) \mathds{1} -\mathds{D}  \right)^{-1}\mathds{R}^{-1} \mathds{I}
\end{align}
Then we can simplify the integrand of Eq.(~\ref{eq:FluctuationDissipation}) in the limit of $\epsilon\rightarrow 0$
\begin{align}
&\lim_{\epsilon\rightarrow 0} G^{R}(\omega)-G^{A}(\omega)=-2\pi\mathds{R}\hat{\delta}(\omega-\mathds{D})\mathds{R}^{-1}\mathds{I}&
\end{align}
where $\hat{\delta}(\omega-\mathds{D})=\text{diag}\big[\omega-\mathds{D}_{1,1},\cdots,\omega-\mathds{D}_{4N,4N}\big]$. This yields the following formula for the correlation functions
\begin{align}
&\hat{C}=\mathds{R}\hat{D}\mathds{R}^{-1}\mathds{I}&,
\end{align}
with $\hat{D}=\text{diag}\left[1+\tanh\left(\dfrac{\beta\mathds{D}_{1,1}}{2}\right),\cdots,1+\tanh\left(\dfrac{\beta\mathds{D}_{4N,4N}}{2}\right)\right]$. In the translation invariant case, we can block-diagonalize $\mathds{G}^{R/A}(\omega)$ in Fourier space which significantly accelerates the computations.

\subsection{Self-consistent equations in terms of correlation function\label{App:Full_SCE}}
We present the expressions for the single-particle self-consistent parameters directly in terms of correlation functions.
\begin{align}
&n_s(i)=2\left(1-C^{11}_0(i)-C^{22}_0(i)-C^{12}_0(i)-C^{21}_0(i)\right)&\\
&n_p(i)=2\left(1-C^{33}_0(i)-C^{44}_0(i)-C^{34}_0(i)-C^{43}_0(i)\right)&\\
&e_s(i)=\sum\limits_{j=i\pm \delta}  \left(C^{11}_1(i,j)-C^{22}_1(i,j)\right)&\\
&e_p(i)=\sum\limits_{j=i\pm \delta} \left(C^{33}_1(i,j)-C^{44}_1(i,j)\right)&\\
e_{sp}(i)=&\sum\limits_{j=i\pm \delta_x} a(i,j) \left(C^{31}_1(i,j)+C^{41}_1(i,j)+C^{23}_1(i,j)+C^{24}_1(i,j)\right)&\\
\nonumber&+\sum\limits_{j=i\pm \delta_y} a(i,j) \left(C^{31}_1(i,j)+C^{41}_1(i,j)-C^{23}_1(i,j)-C^{24}_1(i,j)\right)&\\
e_{ps}(i)=&\sum\limits_{j=i\pm \delta_x} a(i,j) \left(C^{13}_1(i,j)+C^{23}_1(i,j)+C^{41}_1(i,j)+C^{42}_1(i,j)\right)&\\
\nonumber&+\sum\limits_{j=i\pm \delta_y} a(i,j) \left(C^{13}_1(i,j)+C^{23}_1(i,j)-C^{41}_1(i,j)-C^{42}_1(i,j)\right)&
\end{align}

\subsection{Roth decoupling}
\label{Appendix:Roth}
We use the Roth decoupling scheme~\cite{roth1969electron} to compute two-point correlation functions, including density-density, spin-spin, and pair-pair correlations, as described in Eq.(\ref{Eq:pab}). This allows us to express $p$ in terms of both on-site and intersite correlations. The detailed formalism can be found in Refs.\cite{pangburn2024spontaneous,roth1969electron}.

\begin{align}
\label{eq:Roth_DeltaDelta}&\langle\Delta_{i,\alpha}\Delta^\dagger_{j,\beta} \rangle=\dfrac{\rho^\Delta_{\alpha\beta}(i,j)}{1-\phi_{\alpha\beta}(i)}&\\
\label{eq:Roth_SplusSminus}&\langle S^-_{i,\alpha}S^+_{j,\beta} \rangle=-\dfrac{\rho^S_{\alpha\beta}(i,j)}{1+\phi_{\alpha\beta}(i)}&\\
\label{eq:Roth_nn}&\langle n_{i,\alpha\sigma}n_{j,\beta\sigma} \rangle=\dfrac{\left(-1-\phi_{\alpha\beta}(i)+2\left(\rho_{0,\alpha\beta}(i,j)+\phi_{\alpha\beta}(i)(\rho_{0,\alpha\beta}(i,j))+\rho^{n_\sigma n_\sigma} _{\alpha\beta}(i,j)\right)\right)}{2 \left(\phi_{\alpha\beta}(i)\right)^2-2}&
\end{align}

\begin{align}
&\phi_{ss}(i)=\phi_{sp}^\Delta(i)=\dfrac{2}{2-n_s(i)}\left(C_0^{11}(i)+C_0^{21}(i)\right)-\dfrac{2}{n_s(i)}\left(C_0^{12}(i)+C_0^{22}(i)\right)&\\
&\phi_{pp}(i)=\phi_{ps}^\Delta(i)=\dfrac{2}{2-n_p(i)}\left(C_0^{33}(i)+C_0^{43}(i)\right)-\dfrac{2}{n_p(i)}\left(C_0^{34}(i)+C_0^{44}(i)\right)&\\
&\rho_{0,ss}(i)=\dfrac{2}{2-n_s(i)}\dfrac{n_s(i)}{2}\left(C_0^{11}(i)+C_0^{21}(i)\right)&\\
&\rho_{0,pp}(i)=\dfrac{2}{2-n_p(i)}\dfrac{n_p(i)}{2}\left(C_0^{33}(i)+C_0^{43}(i)\right)&\\
&\rho_{0,sp}(i)=\dfrac{2}{2-n_s(i)}\dfrac{n_p(i)}{2}\left(C_0^{11}(i)+C_0^{21}(i)\right)&\\
&\rho_{0,ps}(i)=\dfrac{2}{2-n_p(i)}\dfrac{n_s(i)}{2}\left(C_0^{33}(i)+C_0^{43}(i)\right)&
\end{align}

\begin{align}
&\rho_{ss}^\Delta(i,j)=\dfrac{2}{2-n_s(i)}\left(C_1^{11}+C_1^{21}\right)\left(C_1^{21}+C_1^{22}\right)+\dfrac{2}{n_s(i)}\left(C_1^{11}+C_1^{12}\right)\left(C_1^{12}+C_1^{22}\right)&\\
&\rho_{pp}^\Delta(i,j)=\dfrac{2}{2-n_p(i)}\left(C_1^{33}+C_1^{43}\right)\left(C_1^{43}+C_1^{44}\right)+\dfrac{2}{n_p(i)}\left(C_1^{33}+C_1^{34}\right)\left(C_1^{34}+C_1^{44}\right)&\\
&\rho_{sp}^\Delta(i,j)=\dfrac{2}{2-n_p(i)}\left(C_1^{13}+C_1^{23}\right)\left(C_1^{41}+C_1^{42}\right)+\dfrac{2}{n_p(i)}\left(C_1^{31}+C_1^{32}\right)\left(C_1^{14}+C_1^{24}\right)&\\
&\rho_{ps}^\Delta(i,j)=\dfrac{2}{2-n_s(i)}\left(C_1^{31}+C_1^{41}\right)\left(C_1^{23}+C_1^{24}\right)+\dfrac{2}{n_s(i)}\left(C_1^{13}+C_1^{14}\right)\left(C_1^{32}+C_1^{42}\right)&
\end{align}
\begin{align}
&\rho_{ss}^S(i,j)=\dfrac{2}{2-n_s(i)}\left(C_1^{11}+C_1^{21}\right)\left(C_1^{11}+C_1^{12}\right)+\dfrac{2}{n_s(i)}\left(C_1^{22}+C_1^{12}\right)\left(C_1^{21}+C_1^{22}\right)&\\
&\rho_{pp}^S(i,j)=\dfrac{2}{2-n_p(i)}\left(C_1^{33}+C_1^{43}\right)\left(C_1^{33}+C_1^{34}\right)+\dfrac{2}{n_p(i)}\left(C_1^{44}+C_1^{34}\right)\left(C_1^{43}+C_1^{44}\right)&\\
&\rho_{sp}^S(i,j)=\dfrac{2}{2-n_p(i)}\left(C_{1}^{23}+C_1^{13}\right)\left(C_1^{31}+C_1^{32}\right)+\dfrac{2}{n_p(i)}\left(C_1^{14}+C_1^{24}\right)\left(C_1^{41}+C_1^{42}\right)&\\
&\rho_{ps}^S(i,j)=\dfrac{2}{2-n_s(i)}\left(C_1^{41}+C_1^{31}\right)\left(C_1^{13}+C_1^{14}\right)+\dfrac{2}{n_s(i)}\left(C_1^{32}+C_1^{42}\right)\left(C_1^{23}+C_1^{24}\right)&
\end{align}
\begin{align}
&\rho_{ss}^{n_\sigma n_\sigma}(i,j)=\dfrac{2}{2-n_s(i)}\left(C_1^{11}+C_1^{21}\right)\left(C_1^{11}+C_1^{12}\right)+\dfrac{2}{n_s(i)}\left(C_1^{22}+C_1^{12}\right)\left(C_1^{21}+C_1^{22}\right)&\\
&\rho_{pp}^{n_\sigma n_\sigma}(i,j)=\dfrac{2}{2-n_p(i)}\left(C_1^{33}+C_1^{43}\right)\left(C_1^{33}+C_1^{34}\right)+\dfrac{2}{n_p(i)}\left(C_1^{44}+C_1^{34}\right)\left(C_1^{43}+C_1^{44}\right)&\\
&\rho_{sp}^{n_\sigma n_\sigma}(i,j)=\dfrac{2}{2-n_p(i)}\left(C_1^{13}+C_1^{23}\right)\left(C_1^{31}+C_1^{32}\right)+\dfrac{2}{n_p(i)}\left(C_1^{14}+C_{1}^{24}\right)\left(C_{1}^{41}+C_1^{42}\right)&\\
&\rho_{ps}^{n_\sigma n_\sigma}(i,j)=\dfrac{2}{2-n_s(i)}\left(C_1^{31}+C_1^{41}\right)\left(C_1^{13}+C_1^{14}\right)+\dfrac{2}{n_s(i)}\left(C_1^{32}+C_1^{42}\right)\left(C_1^{23}+C_1^{24}\right)&
\end{align}
Where all the $C^{ab}_1=C^{ab}_1(i,j)$.

\subsection{Uniform system}
\label{App:uniform}
 Thus, the $\mathds{I}$-matrix is diagonal. For the uniform system the diagonal elements are given by
\begin{align}
\mathds{I}=\text{diag}\left[\left(1-n_s/2,n_s/2,1-n_p/2,n_p/2\right)\right],
\end{align}
where we have defined average orbital electron density as $n_\alpha=(1/N)\sum_{i,\sigma} \langle \hat{n}_{i,\alpha,\sigma} \rangle$. 
The uniform and $\mathcal{C}_4$ symmetric $\mathds{M}$-matrix becomes 
\begin{align}
\mathds{M}_{11}(\mathbf{k})&=-(\mu+M)(1-n_s/2)-te_{s}-te_{sp}- \gamma(\mathbf{k})\left(1-n_s+p_{ss}\right)&\\
\mathds{M}_{12}(\mathbf{k})&=te_{s}+te_{sp}-\gamma(\mathbf{k})\left(n_s/2-p_{ss}\right)&\\
\mathds{M}_{22}(\mathbf{k})&=-(\mu-U_s+M)n_s/2-te_{s}-te_{sp}-\gamma(\mathbf{k})p_{ss}&\\
\mathds{M}_{33}(\mathbf{k})&=-(\mu-M)(1-n_p/2)+te_{p}-te_{ps} +\gamma(\mathbf{k})\left(1-n_p+p_{pp}\right)&\\
\mathds{M}_{34}(\mathbf{k})&=-te_{p}+te_{ps}+\gamma(\mathbf{k})\left(n_p/2-p_{pp}\right)&\\
\mathds{M}_{44}(\mathbf{k})&=-(\mu-U_p-M)n_p/2+te_{p}-te_{ps}+\gamma(\mathbf{k})p_{pp}&\\
\mathds{M}_{13}(\mathbf{k})&=\sum_\alpha\Gamma_\alpha(\mathbf{k})\left(1-n_s/2-n_p/2+p_{sp}\right)&\\
\mathds{M}_{14}(\mathbf{k})&=\sum_\alpha\Gamma_\alpha(\mathbf{k})\left(n_p/2-p_{sp}\right)&\\
\mathds{M}_{23}(\mathbf{k})&=\sum_\alpha \Gamma_\alpha(\mathbf{k})\left(n_s/2-p_{sp}\right)&\\ 
\mathds{M}_{24}(\mathbf{k})&=\sum_\alpha \Gamma_\alpha(\mathbf{k}) p_{sp}&\\
\mathds{M}_{31}(\mathbf{k})&=\sum\limits_{\alpha}\left[\Gamma_\alpha(\mathbf{k})\right]^*\left(1-n_s/2-n_p/2+p_{sp}\right)&\\
\mathds{M}_{32}(\mathbf{k})&=\sum\limits_{\alpha}\left[\Gamma_\alpha(\mathbf{k})\right]^*\left(n_s/2-p_{sp}\right)&
\\
\mathds{M}_{41}(\mathbf{k})&=\sum\limits_{\alpha}\left[\Gamma_\alpha(\mathbf{k})\right]^*\left(n_p/2-p_{sp}\right)&\\
\mathds{M}_{42}(\mathbf{k})&=\sum\limits_{\alpha}\left[\Gamma_\alpha(\mathbf{k})\right]^*p_{sp}&
\label{eq:Mmatrix}
\end{align}
where the summation over $\alpha$ is along the $x$ and $y$ direction.
Because the $M$-matrix is hermitian and the $I$-matrix is diagonal, the eigenvalues of the $E$-matrix defined in Eq.~(\ref{Eq:Emat}) are real~\cite{roth1969electron}. Also
\begin{align}
&\gamma(\mathbf{k})=2 t \left(\cos(k_x)+\cos(k_y) \right)&\\
&\Gamma_x(\mathbf{k})=2t\sin(k_x)&\\
&\Gamma_y(\mathbf{k})=2it\sin(k_y)&
\end{align}

Furthermore, we introduced the following nearest neighbor bond-expectation value
\begin{align}
e_{\alpha }&=\frac{1}{8N}\sum_{\langle i,j \rangle,\sigma} \left(\langle\xi_{j \alpha \sigma}\xi^\dagger_{i \alpha\sigma}\rangle-\langle\eta_{j \alpha \sigma}\eta^\dagger_{i \alpha \sigma}\rangle\right)&\\
e_{sp}&=\frac{1}{4N}\sum_{\langle i,j \rangle_{\hat{x}},\sigma}a(i,j)\left(\langle c_{jp\sigma}\xi^\dagger_{is\sigma}\rangle+\langle c^\dagger_{jp\sigma}\eta_{is\sigma}\rangle\right)+\frac{1}{4N}\sum_{\langle i,j \rangle_{\hat{y}},\sigma} a(i,j)\left(\langle c_{jp\sigma}\xi^\dagger_{is\sigma}\rangle-\langle c^\dagger_{jp\sigma}\eta_{is\sigma}\rangle\right)&\\
e_{ps}&=\frac{1}{4N}\sum_{\langle i,j \rangle_{\hat{x}},\sigma}a(i,j)\left(\langle c_{js\sigma}\xi^\dagger_{ip\sigma}\rangle+\langle c^\dagger_{js\sigma}\eta_{ip\sigma}\rangle\right)+\frac{1}{4N}\sum_{\langle i,j\rangle_{\hat{y}},\sigma }a(i,j)\left(\langle c_{js\sigma}\xi^\dagger_{ip\sigma}\rangle-\langle c^\dagger_{js\sigma}\eta_{ip\sigma}\rangle\right)&\\
p_{\alpha  \beta}&=\frac{1}{4N}\sum_{\langle i,j \rangle}\langle \hat{n}_{i \alpha \downarrow}\hat{n}_{j \beta \downarrow}\rangle+\langle S^-_{i \alpha }S^+_{j \beta}\rangle-\langle \Delta_{i \alpha }\Delta^\dagger_{j \beta}\rangle
\label{Eq:pab}
\end{align}
with $S_{i\alpha}^-=c^\dagger_{i\alpha\downarrow}c_{i\alpha\uparrow}$, $S_{i\alpha}^+=c^\dagger_{i\alpha\uparrow}c_{i\alpha\downarrow}$ and $\Delta_{i\alpha}=c_{i\alpha\uparrow}c_{i\alpha\downarrow}$. With, $a(i,i+\hat{x})=i$,$a(i,i-\hat{x})=-i$, $a(i,i+\hat{y})=1$ and $a(i,i-\hat{y})=-1$. Because we are using the paramagnetic assumption, $\langle \hat{n}_{i\alpha\downarrow} \hat{n}_{j\beta\downarrow} \rangle=\langle \hat{n}_{i\alpha\uparrow} \hat{n}_{j\beta\uparrow} \rangle$. These quantitites needs to be evaluated self-consistently. See the Section of the non-homogeneous system for more details.

\subsubsection{Winding number from Composite operator}
Additionally, the topological invariant associated with the composite operator bands by using the composite Green function $\mathds{G}$ of the composite operators and the same winding number formula. Using properties of the winding number~\cite{bollmann2023topological}, we have
\begin{align}
&N_3\big[\mathds{G}\big]=N_3\big[\left(\omega-\mathds{E}\right)^{-1}\big]+N_3\big[\mathds{I}\big]=N_3[\left(\omega-\mathds{E}\right)^{-1}]&
\end{align}
This demonstrates $N_3[\mathds{G}]$ can be computed as if $\mathds{E}$ is the tight-binding Hamiltonian of non-interacting electrons. Because $\eta$ and $\xi$ are weakly interacting quasiparticles in the composite operator formalism, there is no self-energy corrections to the composite operators that modifies the correspondence between the Chern number $\mathcal{C}$ and $N_3\big[\mathds{G}\big]$~\cite{peralta2023connecting}. We checked numerically that if zeros are not taking into account in the computation of $N_3[\mathcal{G}]$, we have $N_3[\mathcal{G}]=N_3[\mathds{G}]$.

\section{Comparing two-point Green's function from composite operator with exact results on the Hubbard Dimer\label{App:Dimer}}
\begin{figure}[h!]
\centering
\includegraphics[width=12.5cm]{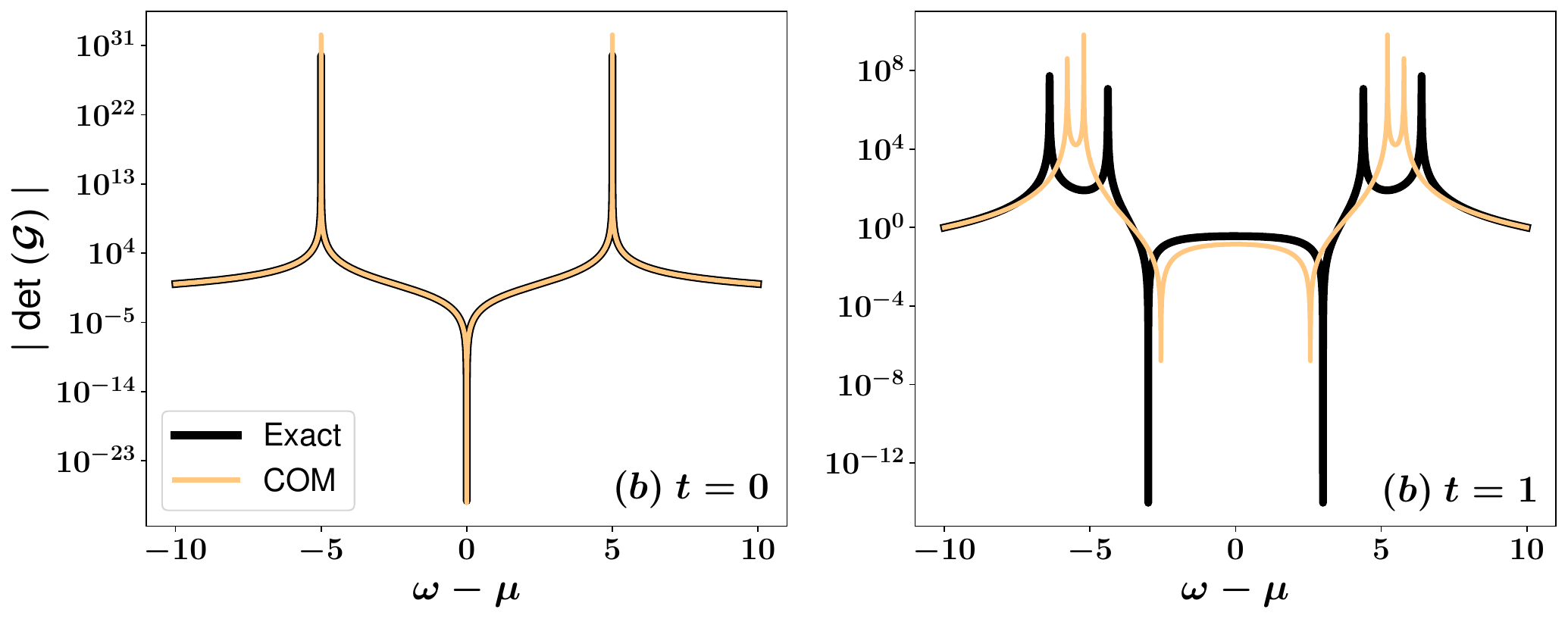}
\caption{Comparison of $|\det \mathcal{G}|$ for the Hubbard dimer between the exact solution and the composite operator method. (a) In the atomic limit ($t=0$), COM exactly reproduces the analytic result, correctly capturing both the positions of poles and zeros. (b) For finite hopping ($t=1$), COM deviates from the exact solution by underestimating the hybridization between the sites. Nevertheless, the overall structure, including the number of poles and zeros, remains qualitatively preserved.}
\label{fig:AppDimer}
\end{figure}
In this Appendix, we benchmark the holon-doublon composite operator method against the exact solution of the single-band Hubbard dimer~\cite{Correl21}. Figure~\ref{fig:AppDimer} shows the absolute value of the determinant of the Green's function, $|\det(\mathcal{G})|$, for $U=10$. The maxima of $|\det(\mathcal{G})|$ indicate the positions of the poles, while the minima correspond to the positions of the zeros.

The composite operator method becomes exact in the atomic limit ($t = 0$), shown in Fig.~\ref{fig:AppDimer} (a). Accordingly, it reproduces the analytic results presented in Ref.~\cite{Correl21}. The pole and zero positions are matched precisely in the atomic limit.

However, the method is no longer exact when a finite hopping is introduced ($t \neq 0$), as illustrated in Fig.~\ref{fig:AppDimer} (b). The composite operator approach underestimates the hybridization between the two sites. This behavior is expected because although the interactions are treated exactly within this formalism  yet the hopping is incorporated via a truncated equation of motion for the Green's function. Nevertheless, in the limit $U \gg t$, the number of poles and zeros remains preserved, even though their precise positions may shift.

Our primary focus is on the topological character of the zeros and poles, rather than their exact dispersions. As long as the spectral gap remains open, these topological features are robust against small quantitative perturbations. Consequently, the Hubbard operator method offers an efficient and reliable framework for investigating topological properties. While its accuracy can be further enhanced by incorporating higher-order correlations, such improvements come at the cost of increased computational complexity.

\section{Junction between TMZ/TMBI phases and trivial Mott insulator\label{App:Junction}}
This section examines various types of junctions to test the bulk-boundary correspondence associated with the single-particle winding number $N^\sigma_3$ and the Chern number $\mathcal{C}$. Figure~\ref{fig:fig6App} presents the minimum and maximum eigenvalues of the Green’s function across two distinct junction geometries.

The top panels of Fig.~\ref{fig:fig6App} correspond to a junction between a TMZ phase with $N^\sigma_3 \ne 0$ and $\mathcal{C}_\sigma = 0$, and a trivial Mott insulator with $N^\sigma_3 = \mathcal{C}_\sigma = 0$. The bottom panels show a junction between a topological Mott band insulator (TMBI) phase characterized by $N^\sigma_3 = \mathcal{C}_\sigma \ne 0$ and a trivial Mott insulator.

In the TMBI–MI junction, in-gap edge states appear as expected, while the Green’s function zeros remain gapped. In contrast, the TMZ–MI junction exhibits in-gap zero modes but no edge states that can carry current. These observations confirm the bulk-boundary correspondence for both topological invariants: the single-particle winding number $N^\sigma_3$, associated with zeros of the Green’s function, and the many-body Chern number $\mathcal{C}_\sigma$.

\begin{figure}[h!]
\centering
\includegraphics[width=12.5cm]{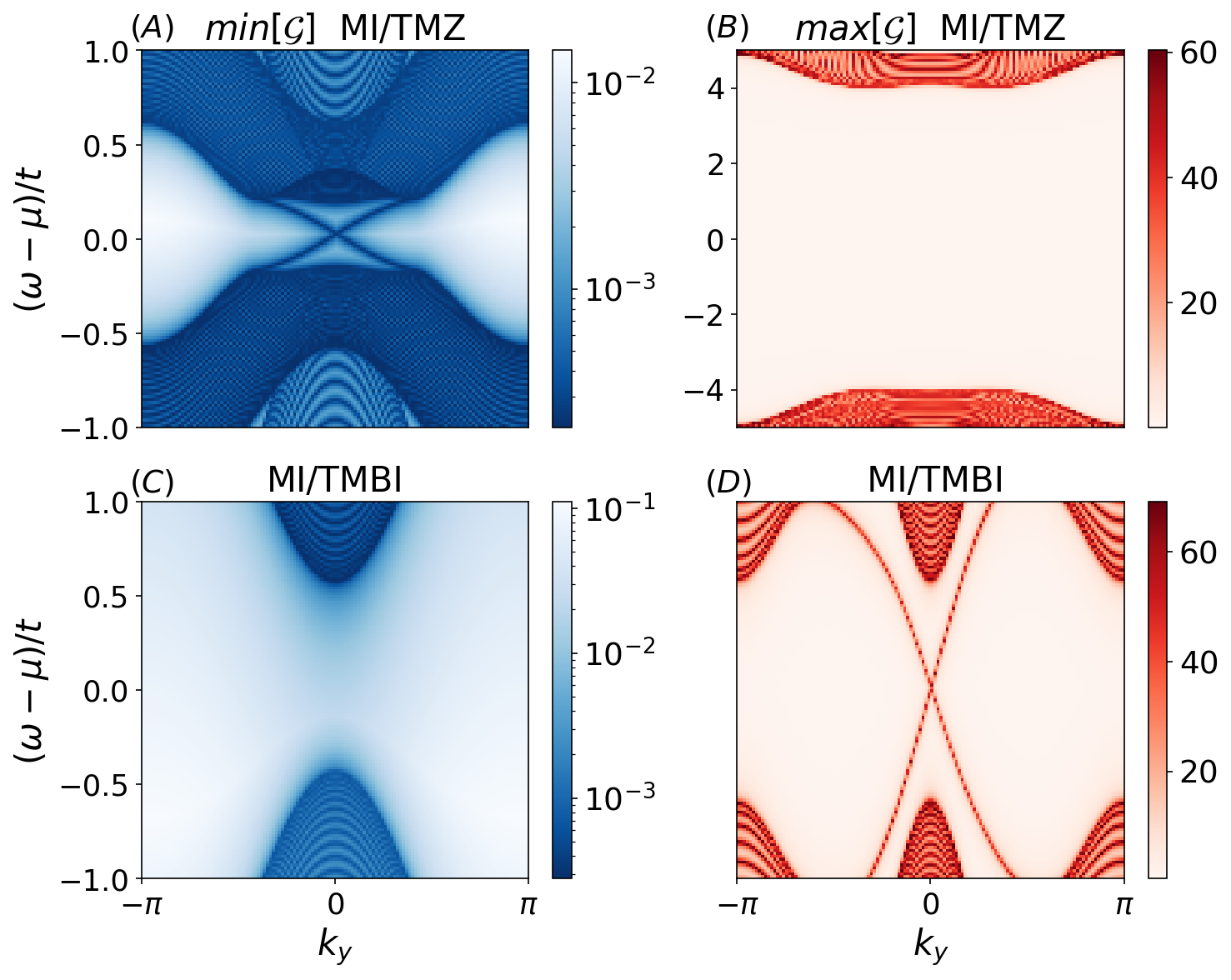}
\caption{The minimum (left column) and maximum (right column) eigenvalue of the single particle Green function $\mathcal{G}$ are shown in function of $\omega$ and $k_y$. Both (A/B) and (C/D) corresponds to the same single-particle Green function $\mathcal{G}$. In both cases, we select a representative set of self-consistent parameters for a system divided into two halves. One half is in the MI phase with $N_{3,\sigma}=\mathcal{C}=0$, while the other half is in the TMZ phase with $N_{3,\sigma}=-1$ in the upper row (A/B) and and in the TMBI phase with $N_{3,\sigma}=\mathcal{C}_\sigma=1$ in the lower row (C/D).}
\label{fig:fig6App}
\end{figure}

\section{Calculation of Hall conductivity and St\v{r}eda formula}
\label{App:Streda}
\begin{figure}[h!]
\centering
\includegraphics[width=12.5cm]{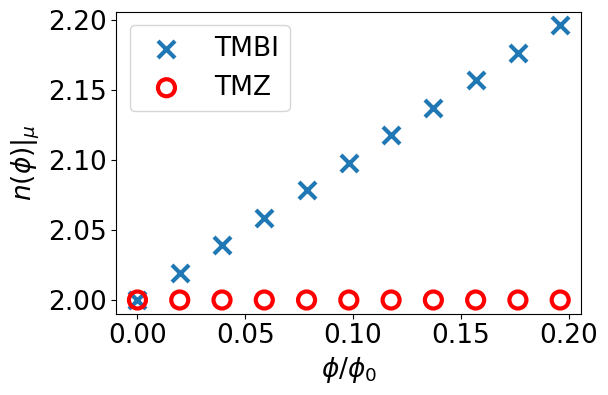}
\caption{Variation of the electron density $n(\phi)|_{\mu}$ in function of the magnetic flux $\phi$. The blue cross ($x$) show the variation of $n$ in the TMBI phase, where the winding number and Chern number satisfy $N_{3,\sigma} = \mathcal{C}_\sigma =1$. The red circle ($o$) shows the variation of $n$ in the TMZ phase, where $N{3,\sigma} \ne 0$ but the Chern number remains trivial $\mathcal{C}_\sigma = 0$.}
\label{fig:fig8}
\end{figure}

The many-body Chern number $\mathcal{C}_\sigma$ of a gapped interacting system can be computed using the St\v{r}eda formula~\cite{streda1982theory}, which relates the Hall conductivity to the change in electronic density with applied magnetic field

\begin{align}
\sigma_{xy} = \phi_0 \left.\frac{\partial n}{\partial B}\right|_{\mu, T = 0},
\label{eq:Streda_formula}
\end{align}
where $\phi_0$ is the magnetic flux quantum. The electron density $n$ can be expressed in terms of the Green’s function as:

\begin{align}
n(\phi) = \frac{1}{\beta \Omega} \sum_{n} e^{z_n 0^+} \, \text{Tr} \left[ \mathcal{G}^\phi(z_n) \right],
\end{align}

with $z_n = i\omega_n = i(2n + 1)\pi / \beta$ the fermionic Matsubara frequencies, $\beta$ the inverse temperature, and $\Omega$ the unit cell area. We use this approach throughout the manuscript to verify the many-body Chern number.

Recent studies~\cite{peralta2023connecting} have emphasized that applying a magnetic field reveals a fundamental distinction between the many-body Chern number $\mathcal{C}_\sigma$ and the single-particle winding number $N_{3,\sigma}$, particularly in systems that are not adiabatically connected to a non-interacting limit.

To incorporate magnetic fields, we adopt the standard Peierls substitution~\cite{hofstadter1976energy,herzog2020hofstadter}. For a path $\gamma_{\mathbf{r} \rightarrow \mathbf{r}'}$ connecting two lattice sites $\mathbf{r}$ and $\mathbf{r}'$, the Peierls phase is:

\begin{align}
\phi_{\mathbf{rr'}} = \int_{\gamma_{\mathbf{r} \rightarrow \mathbf{r}'}} \mathbf{A} \cdot d\mathbf{l},
\end{align}

where $\mathbf{A}$ is the electromagnetic vector potential. The hopping amplitudes are then modified as:
\begin{align}
t_{\alpha\beta}(\mathbf{r} - \mathbf{r}') \rightarrow e^{i \phi_{\mathbf{rr'}}} t_{\alpha\beta}(\mathbf{r} - \mathbf{r}').
\end{align}

We choose the Landau gauge, $\mathbf{A} = (-\phi y, 0)$, where $\phi$ is the magnetic flux per plaquette. For generic $\phi$, translational symmetry is broken, but it is restored for rational values $\phi = 2\pi p/q$ with $p$ and $q$ coprime. In such cases, a magnetic unit cell of size $1 \times q$ can be defined, enabling access to small magnetic fields by choosing sufficiently large $q$.

As a benchmark, the spinless non-interacting Chern insulator model under flux $\phi = 2\pi p/q$ can be written as:

\begin{align}
\mathcal{H}_{y, y'}^{\phi} &= \delta_{y, y'} h_y + \delta_{y, y'+1} T + \delta_{y+1, y'} T^\dagger, \\
h_y &= \left(M - 2 \cos(k_x - \phi y)\right) \tau_z + 2 \sin(k_x - \phi y) \tau_x, \\
T &= e^{-i k_y} \tau_y,
\end{align}

where $y, y' \in \{0, 1, \dots, q - 1\}$. To include interaction effects, we apply the composite operator method (COM) to the modified hopping Hamiltonian $\mathcal{H}^\phi$ incorporating the magnetic field.

Figure~\ref{fig:fig8} shows the dependence of the electronic density $n$ on the magnetic field $B$ at fixed chemical potential $\mu$. The slope of $n$ versus $B$ allows us to extract the many-body Chern number via Eq.~\eqref{eq:Streda_formula}. As expected, we find that the topological Mott band insulator (TMBI) phase exhibits a quantized Chern number $\mathcal{C}_\sigma = 1$, while the topological Mott zero (TMZ) phase yields $\mathcal{C}_\sigma = 0$.

\section{Static $\pi$-flux defect}
\label{App:Pi}
\begin{figure}[h!]
\centering
\includegraphics[width=12.5cm]{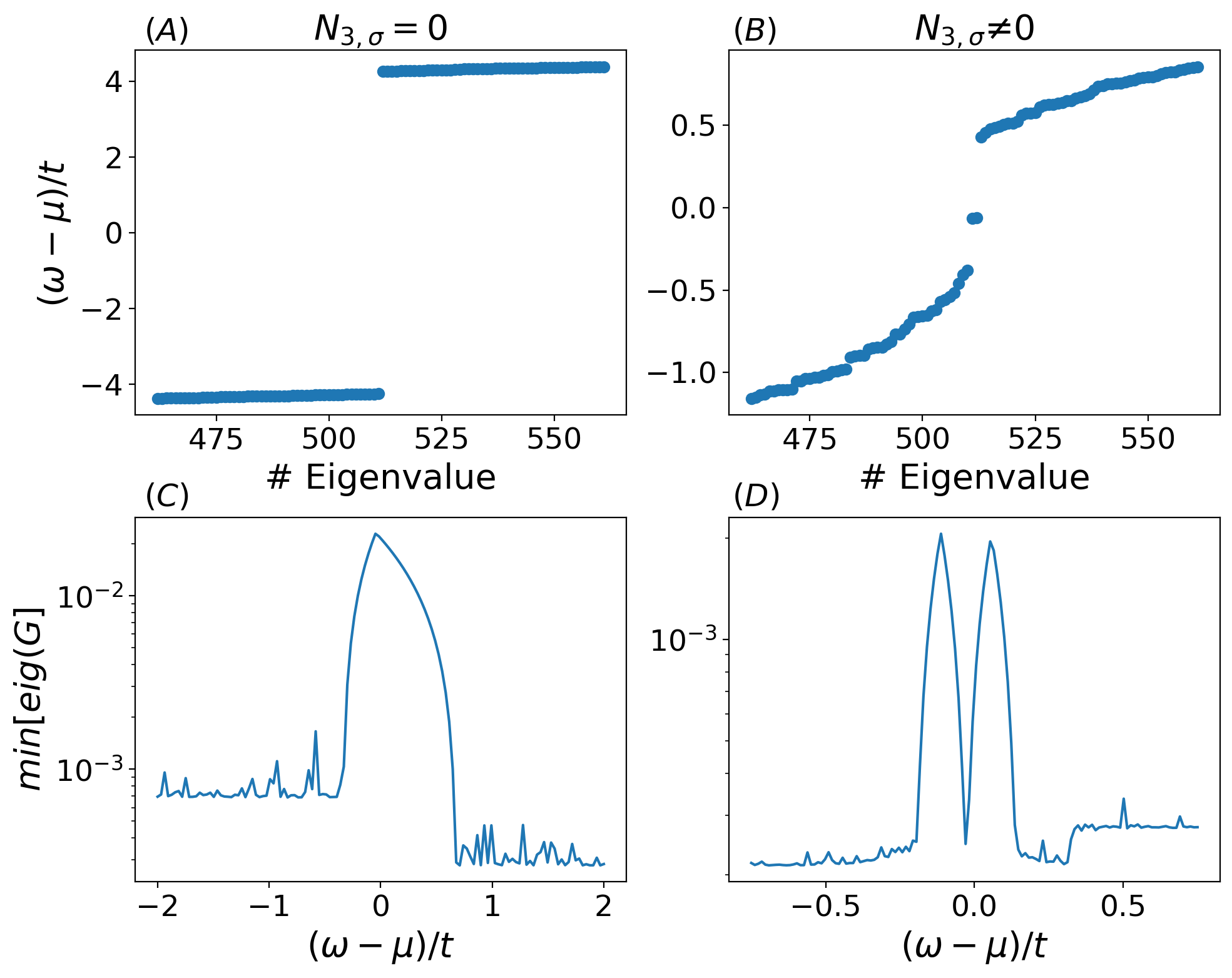}
\caption{Response to a $\pi$-flux defect for the Mott insulators (A),  $U_s=12.2 t$ $M=-0.28 t$, (C), $M=0.53 t$ $U_s=7.6t$, , TMBI (B), $U_s=12.2t$ $M=4.36$ and TMZ (D) phase. The upper row (A/B) shows the $E$-matrix eigenvalues with the insertion of two $\pi$- flux with periodic boundary conditions (PBC). The lower row (C/D) displays $min\left[eig(G)\right]$ in function of $\omega$ with the insertion of two $\pi$-flux with periodic boundary conditions. 
As expected from the bulk-boundary correspondence the TMBI phase spectrum (B) exhibits a pair of subgap states when a $\pi$-flux is inserted. In contrast, the spectrum of the Mott insulator phase (A) does not exhibit subgap states when a $\pi$-flux is introduced.  Similarly as expected from the bulk-boundary correspondence of zeros, the TMZ phase (D) exhibits a pair of subgap states in the $\pi$-flux case. In contrast, the zeros spectrum (C) does not exhibit zero-subgap states when a $\pi$-flux is introduced.}
\label{fig:fig7}
\end{figure}
In non-interacting systems, local topological defects such as static $\pi$-flux~\cite{schindler2022topological}, disclinations or dislocations~\cite{ran2008spin,jurivcic2012universal} can bind topological edge states thus acting as a local probe of the underlying topology. In Fig.~(\ref{fig:fig7} A) and Fig.~(\ref{fig:fig7} B), we show the excitation spectrum in the Mott insulator phase ($N_3=0$) and the TMBI phase ($N_3=-1$), respectively. Subgap states localized at the $\pi$-flux core appear in the topologically non-trivial phase, whereas in the regime where $N_3=0$, there are no subgap states. We verified across the phase diagram that a subgap state was present only when the winding number was non-zero. Thus, bulk boundary correspondence operates like topological insulators, even in the presence of interactions.

Similar to the poles we examine the zeros spectrum upon insertion of a static $\pi$-flux. In both the Mott insulator and TMZ phases, no states corresponding to Green's function poles appear within the single-particle gap. The minimum of the eigenvalues for the trivial MI phase and TMZ phase are shown respectively in Fig.~(\ref{fig:fig7} C) and Fig.~(\ref{fig:fig7} D). In both cases, the two low-value plateaus correspond to the zeros bands. A finite quasiparticle damping is applied, which accounts for the non-zero minimum value $min\left[eig(G)\right]$. Near $\omega=0$, the value of $min\left[(G)\right](\omega)$ is higher, indicating the location of the zeros gap. The key distinction is the presence of a zero mode at $\omega=0$ in the TMZ phase, which is absent in the trivial MI phase.

We can conclude that there is a one-to-one correspondence between a non-zero value of $N_3$ for both Green's function poles and zeros, in both open boundary conditions and in the presence of topological defects.

\section{Choosen set of parameters\label{App:Params}}

In this appendix, we present the parameters used to obtain the results shown in the main text for the junction for the Mott insulator phase (MI), the topological Mott band insulator phase (TMBI) with $C_{\sigma}=1$, the topological Mott zeros phase (TMZ) with $N_{3,\sigma}=-1$ and the topological Mott zeros phase (TMZ) with $N_{3,\sigma}= \pm 1$.

\begin{table}[h!]
\centering
\begin{tabular}{|c|c|c|c|c|c|c|c|c|c|c|c|c|c|}
\hline
Phase & $\mu$ & $n_s$ & $n_p$ & $e_s$ & $e_p$ & $p_s$ & $p_p$ & $e_{sp}$ & $e_{ps}$ & $p_{sp}$ & $M$ & $U_s$ & $U_p$ \\ 
\hline 
MI &  6.04 &  1.00  &      1.00  &      0.10 & 0.08  & 0.10 & 0.12 & -0.27 & -0.27  &0.28 & -0.28 & 12.18 & 12.18 \\ 
\hline
TMBI $C_\sigma=1$& 6.17 & 1.14& 0.86 & 0.05  & 0.08 & 0.24 & 0.01 & -0.48 & -0.48  &0.25 & 4.36& 12.18 & 12.18  \\ 
\hline
TMZ $N_{3,\sigma}=-1$&  5.94 & 1.00  & 1.00  & 0.003 & -0.05 & 0.19  & 0.19 & -0.23 & -0.23 &  0.27 & 0.53 & 13.71 & 12.0 \\ 
\hline
TMZ $N_{3,\sigma}=1$& 5.88 & 1.00  & 1.00 & 0.005 & -0.06  & 0.19 &  0.20 & -0.28 & -0.28 & 0.27 & -0.48 & 13.71 & 12.0 \\ 
\hline
\end{tabular}
\caption{The set of parameters used in the Chern-Hubbard model for the results presented in the main text. The first row corresponds to the trivial Mott insulator (MI) phase, the second row to the topological Mott band insulator (TMBI) phase, the third row to the topological Mott zeros (TMZ) phase with $N_{3,\sigma}=-1$ and the fourth row to the topological Mott zeros (TMZ) phase with $N_{3,\sigma}=1$.}
\label{Table:Params_Chern}
\end{table}

\end{onecolumngrid}
\end{document}